\documentclass[aps,prc,10pt,twocolumn,superscriptaddress,showpacs]{revtex4-1}
\usepackage{graphicx}
\usepackage{color}
\usepackage{bm}
\usepackage{amsmath}
\usepackage{multirow}

\begin{document}

\title{First-principles interatomic potentials for ten elemental metals via compressed sensing}
\author{Atsuto \surname{Seko}}
\email{seko@cms.mtl.kyoto-u.ac.jp}
\affiliation{Department of Materials Science and Engineering, Kyoto University, Kyoto 606-8501, Japan}
\affiliation{Center for Elements Strategy Initiative for Structure Materials (ESISM), Kyoto University, Kyoto 606-8501, Japan}
\author{Akira \surname{Takahashi}}
\affiliation{Department of Materials Science and Engineering, Kyoto University, Kyoto 606-8501, Japan}
\author{Isao \surname{Tanaka}}
\affiliation{Department of Materials Science and Engineering, Kyoto University, Kyoto 606-8501, Japan}
\affiliation{Center for Elements Strategy Initiative for Structure Materials (ESISM), Kyoto University, Kyoto 606-8501, Japan}
\affiliation{Nanostructures Research Laboratory, Japan Fine Ceramics Center, Nagoya 456-8587, Japan}

\date{\today}
\pacs{71.15.Pd,31.50.Bc,34.20.-b,65.40.-b}

\begin{abstract}
Interatomic potentials have been widely used in atomistic simulations such as molecular dynamics.
Recently, frameworks to construct accurate interatomic potentials that combine a systematic set of density functional theory (DFT) calculations with machine learning techniques have been proposed.
One of these methods is to use compressed sensing to derive a sparse representation for the interatomic potential. 
This facilitates the control of the accuracy of interatomic potentials.
In this study, we demonstrate the applicability of compressed sensing to deriving the interatomic potential of ten elemental metals, namely, Ag, Al, Au, Ca, Cu, Ga, In, K, Li and Zn.
For each elemental metal, the interatomic potential is obtained from DFT calculations using elastic net regression.
The interatomic potentials are found to have prediction errors of less than 3.5 meV/atom, 0.03 eV/\AA\ and 0.15 GPa for the energy, force and the stress tensor, respectively, which enable the accurate prediction of physical properties such as lattice constants and the phonon dispersion relationship.
\end{abstract}

\maketitle
\section{Introduction}
Molecular dynamics (MD) has been a popular tool for modeling a collection of interacting atoms within classical mechanics\cite{alder1959studies}.
The relationship between energy and atomic coordinates, namely, the potential energy surface (PES), plays a key role in MD simulations since the PES determines the forces acting on atoms that originate from atomic interactions and therefore the motion of atoms.
As an alternative to first principles MD calculations, which provide the most accurate PES\cite{car1985unified}, frameworks to estimate a reliable PES based on the combination of systematic density functional theory (DFT) calculations and machine learning techniques have recently been proposed, which are applicable to periodic systems\cite{behler2007generalized,bartok2010gaussian}.
The starting point of these methods is that a DFT calculation is performed for at least $10^3$ different atomic configurations.
A PES is then constructed from the DFT training data set using regression techniques to estimate the relationship between predictor and observation variables.
Its accuracy is known to be much better than that of conventional interatomic potentials owing to the flexibility of the method. 
The flexibility also makes it possible to construct the PES for a wide range of materials using the same method.

To estimate the PES from a data set of the energy for many atomic configurations, a variety of methods can be applied.
For applications to molecules and clusters, 
spline methods\cite{Bowman1986,Chapman198393}, 
interpolating moving least-squares methods\cite{Dawes2007_1.2730798,Dowes2008_1.2831790}, 
modified Shepard interpolation and other interpolation techniques\cite{Ischtwan1994,Jordan1995_1.469296,Fournier2013_1.4846297}, 
artificial neural networks\cite{Lorenz2004210,Gassuner1998_jp972209d,Prudente1998_1.477550,Prudente1998585,Brown1996_1.472596,Blank1995_1.469597,Sumpter1992455,Behler2007_1.2746232,Jun2014_1.4832697,Bin2014_1.4887363,Handley2010_jp9105585,Agrawal2009_jp8085232,Nguyen2012_jp3020386,Le2010_jp907507z,Darley2008_ct800166r,Cho200277,Malshe2009_1.3231686,Manzhos2006_1.2336223}
and the reproducing kernel Hilbert space method\cite{Hollebeek1999.50.1.537,Ho2003_1.1603219}
have been used.
However, only a few nonlinear regression techniques such as 
artificial neural networks\cite{behler2007generalized,PhysRevB.81.184107,PhysRevB.83.153101,PhysRevB.85.045439,PhysRevB.85.174103,behler2011neural,behler2014representing} 
and Gaussian process regression\cite{bartok2010gaussian} 
have been adopted to estimate the PES for solids on account of the complex relationship between the energy and crystal structure in solids.
Thus, applications to solids have been limited to a small number of metallic\cite{PhysRevB.81.184107,PhysRevB.85.045439}, covalent\cite{behler2007generalized,bartok2010gaussian,PhysRevB.85.174103} and ionic materials\cite{PhysRevB.83.153101}.

In these methods, the PES is generally estimated by transforming atomic positions into some descriptors. 
This plays an essential role in constructing a PES that satisfies several invariances, such as translational and rotational invariance, and is transferable to structures composed of a different number of atoms from those in the training data.
Obviously, the accuracy of the PES strongly depends on the selection of the descriptors.
So far, several descriptors for expressing atomic coordinates have been proposed\cite{behler2011atom,jose2012construction,bartok2013representing,PhysRevLett.108.058301,PhysRevB.89.205118,von2013representation}, although only some of them have succeeded in obtaining accurate PESs.
Descriptors are preferably chosen without a priori knowledge of the energetics of the target material.
Therefore, it is desirable to establish a method that enables automatic optimization of the descriptors for constructing a PES.

The use of least absolute shrinkage and selection operator (LASSO) regression\cite{tibshirani1996regression,hastieelements} is a promising means of enabling the automatic selection of descriptors. 
We previously introduced a simple scheme to estimate the PES using LASSO regression that was based on a set of simple and systematic basis functions and a linear relationship between the energy and basis functions\cite{PhysRevB.90.024101}, which was applied to the elemental metals of Na and Mg.
It was found that a sparse representation for the PES with a small number of basis functions was efficiently derived from relatively a large number of systematic candidate basis functions depending only on the distances between atoms.
We also found that the energy can be expressed by a linear relationship with the basis functions.
As a result, sparse PESs with prediction errors of 1.3 and 0.9 meV/atom were obtained for Na and Mg, respectively.
In addition to our application of LASSO regression for PES construction, it has recently been used to obtain sparse representations for alloy thermodynamics\cite{nelson2013compressive}, interatomic force constants\cite{PhysRevLett.113.185501} and the PES for a molecule\cite{Mizukami2014_1.4897486}.

The scheme to construct the PES on the basis of LASSO regression with a linear relationship for energy has a number of advantages:
1) Accuracy can be controlled in a transparent manner.
2) A well-optimized sparse representation for the PES is obtained, which can accelerate and increase the accuracy of atomistic simulations while decreasing the computational costs.
3) Information on the forces acting on atoms and stress tensors can be included in the training data in a straightforward manner.
4) Regression coefficients are generally determined quickly using a standard least-squares technique.
5) The number of regression coefficients does not explicitly depend on the size of the input data set.

In this study, we apply this scheme to construct PESs for ten elemental metals, namely, Ag, Au, Al, Ca, Cu, Li, K, Ga, In and Zn.
Here, we use elastic net regression, which is a generalization of LASSO regression.
This paper is organized as follows.
Sec. \ref{lasso2:methodology_section} presents the methodology including the linear expression for the total energy, the systematic basis functions and the regression techniques.
Linear expressions for the forces acting on atoms and the stress tensors are also derived from the expression for the total energy.
In Sec. \ref{lasso2:preparation_input_data_section}, the procedure for optimizing the input factors used to estimate the PES is described.
In Sec. \ref{lasso2:elastic_net_section}, the application of elastic net regression to the ten elemental metals is demonstrated.
Finally, we give a conclusion in Sec. \ref{lasso2:conclusion_section}.

\section{Methodology}
\label{lasso2:methodology_section}
\subsection{Linear expressions for total energy, forces acting on atoms and stress tensor}
\begin{figure}[tbp]
\begin{center}
\includegraphics[width=\linewidth,clip]{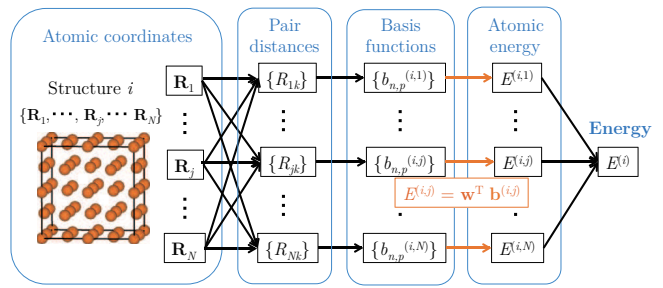} 
\caption{
Relationship between total energy and crystal structure.
}
\label{lasso2:linear_model}
\end{center}
\end{figure}

To model the relationship between the total energy and the crystal structure, we adopt the linear expression for the total energy proposed in Ref. \onlinecite{PhysRevB.90.024101}.
Figure \ref{lasso2:linear_model} shows the linear model for the total energy, which is based on the widely accepted idea that the total energy of a structure is equal to the sum of its atomic energies\cite{behler2007generalized,bartok2010gaussian}.
Introducing a basis expansion derived from a set of other atomic positions, this model assumes a linear relationship between the energy of atom $j$ in structure $i$, $E^{(i,j)}$, and the set of basis functions for atom $j$ in structure $i$, $\{ b_n ^{(i,j)} \}$.
The linear relationship between the atomic energy and $M$ given basis functions is expressed as
\begin{equation}
E^{(i,j)} = \sum_{n=0}^{M} w_n b^{(i,j)}_n,
\end{equation}
where $w_n$ and $b^{(i,j)}_n$ denote the expansion coefficients and basis functions for atom $j$ of structure $i$, respectively, and $b^{(i,j)}_0=1$.
By applying the same expansion coefficients to identical atomic species, the total energy of structure $i$ composed of $N^{(i)}$ atoms is derived as
\begin{eqnarray}
E^{(i)} & = &\sum_{j} E^{(i,j)} \nonumber \\
    & = & \sum_n w_n \left[ \sum_{j} b^{(i,j)}_n \right] \nonumber \\
    & = & \sum_n w_n x_n^{(i)}, 
\label{lasso2:energy_equation}
\end{eqnarray}
where $x_n^{(i)}$ satisfies
\begin{equation}
x^{(i)}_n = \sum_{j} b^{(i,j)}_n.
\end{equation}
Consequently, the total energy of structure $i$ is expressed as a linear relationship with the sum of the basis functions for all atoms in structure $i$.

The forces acting on atoms and the stress tensor can be given by linear equations as well as the total energy (Appendix \ref{lasso2:force-stress-section} for details).
$\alpha$th component of the force acting on atom $l$ and the virial stress tensor $\sigma_{\alpha\beta}$ of structure $i$ are expressed as
\begin{equation}
F^{(i)}_{l,\alpha} = \sum_n w_n x_{{\rm force},n}^{(i,l,\alpha)}
\label{lasso2:force-linear-model}
\end{equation}
and
\begin{equation}
\sigma_{\alpha\beta}^{(i)} = \sum_n w_n x_{{\rm stress},n}^{(i,\alpha,\beta)},
\end{equation}
respectively, where $x_{{\rm force},n}^{(i,l,\alpha)}$ and $x_{{\rm stress},n}^{(i,\alpha,\beta)}$ can be derived from the derivative of the basis functions with respect to the atomic coordinates as will be shown later.

\subsection{Estimation of regression coefficients}
The expansion coefficients $\bm{w} = [w_0, w_1, \cdots, w_M]^\top$ characterizing the energetics of a system can be estimated by regression, which is a machine learning method for estimating the relationship between the predictor and observation variables using a training data set.
Regarding the training data, the energy, the forces acting on atoms and the stress tensor computed by DFT calculation can all be used as observations in the regression process, since all of them can be expressed by linear equations with the same expansion coefficients.
When considering only the energy as observations, the predictor matrix $\bm{X}$ and observation vector $\bm{y}$ correspond to $\bm{X}_{\rm energy}$ and $\bm{y}_{\rm energy}$, which are composed of $x_n^{(i)}$ and the DFT energies for the structures in the training data, respectively, that is,
\begin{equation}
\bm{X} = \bm{X}_{\rm energy} ,\qquad \bm{y} = \bm{y}_{\rm energy}. 
\end{equation}
When using the energy, forces and stress tensor as observations, $\bm{X}$ and $\bm{y}$ are written as
\begin{equation}
\bm{X} = 
\begin{bmatrix}
\bm{X}_{\rm energy} \\
\bm{X}_{\rm force} \\
\bm{X}_{\rm stress} 
\end{bmatrix}
,\qquad \bm{y} = 
\begin{bmatrix}
\bm{y}_{\rm energy} \\
\bm{y}_{\rm force} \\
\bm{y}_{\rm stress} 
\end{bmatrix}
,
\end{equation}
where $\bm{X}_{\rm force}$ and $\bm{X}_{\rm stress}$ are composed of $x_{{\rm force},n}^{(i,l,\alpha)}$ and $x_{{\rm stress},n}^{(i,\alpha,\beta)}$ for the structures in the training data, respectively.
$\bm{y}_{\rm force}$ and $\bm{y}_{\rm stress}$ are composed of the forces and stress tensor computed by the DFT calculation for the structures in the training data, respectively.


A simple procedure to estimate the expansion coefficients is to use linear ridge regression.
This is a shrinkage method and shrinks the regression coefficients by imposing a penalty.
The ridge coefficients minimize the penalized residual sum of squares expressed as
\begin{equation}
L(\bm{w}) = ||\bm{X}\bm{w} - \bm{y}||_2^2 + \lambda ||\bm{w}||_2^2,
\end{equation}
where $\lambda$ controls the magnitude of the penalty.
This is referred to as $L_2$ regularization. 
The solution is easily obtained only in terms of matrix operations as 
\begin{math}
\bm{w} = (\bm{X}^\top \bm{X} + \lambda \bm{I})^{-1} \bm{X}^\top \bm{y},
\end{math}
where $\bm{I}$ denotes the unit matrix.
Therefore, the regression coefficients can be easily estimated while avoiding the well-known multicollinearity problem occurring in the ordinary least-squares method.

Although linear ridge regression is useful for obtaining a PES from a given basis set, a basis set appropriate for the system of interest is generally unknown.
Moreover, a PES with a small number of basis functions is desirable to decrease the computational cost in atomistic simulations.
Therefore, we use elastic net regression\cite{RSSB:RSSB503,hastieelements} in combination with the preparation of a considerable number of basis functions.
Elastic net regression is a generalization of LASSO regression\cite{tibshirani1996regression,hastieelements} and combines the $L_1$ and $L_2$ penalties.
Elastic net regression enables us not only to provide a solution for linear regression but also to obtain a sparse representation with a small number of nonzero regression coefficients.
The solution is obtained by minimizing the function 
\begin{equation}
L(\bm{w}) = ||\bm{X}\bm{w} - \bm{y}||_2^2 + \alpha \lambda ||\bm{w}||_1 + \frac{(1-\alpha)}{2} \lambda ||\bm{w}||_2^2,
\label{lasso2:elastic_net_min_equation}
\end{equation}
where the parameter $\alpha$ determines the mixing of the penalties.
When $\alpha = 1$, the minimization function corresponds to that of the LASSO.
The accuracy of the solution can be controlled simply by adjusting the values of $\lambda$ and $\alpha$ for a given training data set.

The use of elastic net regression allows us to avoid several limitations of the LASSO.
For example, if there is a group of highly correlated predictor variables, the LASSO tends to select only one variable from the group.
Also, in the case of high-dimensional predictor variables with few observations, the LASSO selects at most the same number of predictor variables as the number of observations before the solution saturates.

Note that the units for the energy, forces and stress tensor are different, hence care is required in the selection of the units.
The units act as weights in the regression.
Here, we used the units of eV/supercell, eV/\AA\ and GPa for the energy, forces and stress tensor, respectively, when considering all of them as observations.
When considering only the energy as observations, the unit of eV/atom can also be used.
In this study, regression coefficients were estimated using the standardized training data.

\subsection{Basis functions}
In this study, the following simple form of the basis functions is newly used as the linear expression for the energy.
The $p$th power of the $n$th element for atom $j$ of structure $i$, $ b_{n,p}^{(i,j)}$, is written as
\begin{equation}
b_{n,p}^{(i,j)} = \left[\sum_k f_n(R_{jk}^{(i)}) \cdot f_c(R_{jk}^{(i)})\right]^p, 
\label{lasso2:basis_function}
\end{equation}
where $p$ is a positive integer and $R_{jk}^{(i)}$ denotes the distance between atoms $j$ and $k$ of structure $i$.
The sum is taken over all atoms within a cutoff radius $R_c$ from atom $j$.
$f_n(R_{jk}^{(i)})$ and $f_c(R_{jk}^{(i)})$ are an analytical pairwise function and a smooth pairwise cutoff function that is zero at a distance greater than $R_c$, respectively.
Since the product of $f_n$ and $f_c$ is pairwise, an exponential form of the sum of the pairwise functions is introduced to take many-body effects into account.
Although pairwise functions are adopted here for $f_n$, other types of basis functions such as angular basis functions can be used in principle.

Taking the sum of the basis functions for all atoms, $x^{(i)}_{n,p}$ is obtained as
\begin{equation}
x_{n,p}^{(i)} = \sum_j \left[\sum_k f_n(R_{jk}^{(i)}) \cdot f_c(R_{jk}^{(i)})\right]^p.
\end{equation}
Using a combination of the linear model and this form of the basis functions, the expression for the total energy is invariant to the translation, rotation and exchange of atoms.
In addition, it can be used to input crystal structures with a different number of atoms from the structures in the training data.

$x_{{\rm force},n,p}^{(i,l,\alpha)}$ and $x_{{\rm stress},n,p}^{(i,\alpha,\beta)}$ can also be derived as
\begin{equation}
x_{{\rm force},n,p}^{(i,l,\alpha)} = - p \sum_j b^{(i,j)}_{n,p-1} \frac{\partial b^{(i,j)}_{n,1}}{\partial  R_{l,\alpha}^{(i)}}
\end{equation}
and
\begin{equation}
x_{{\rm stress},n,p}^{(i,\alpha,\beta)} = - \frac{p}{V} \sum_l R_{l,\alpha}^{(i)} \sum_j b^{(i,j)}_{n,p-1} \frac{\partial b^{(i,j)}_{n,1}}{\partial  R_{l,\beta}^{(i)}}.
\end{equation}
respectively (Appendix \ref{lasso2:force-stress-section} for details).
The derivative of the basis functions with respect to the $\alpha$th component of the atomic position is written as
\begin{equation}
\frac{\partial b^{(j)}_{n,1}}{\partial  R_{l,\alpha}} = \sum_k \left[f_n'(R_{jk}) f_c(R_{jk}) + f_n (R_{jk}) f_c' (R_{jk}) \right] \frac{\partial R_{jk}}{\partial  R_{l,\alpha}}, 
\end{equation}
where the structure index $i$ is omitted.

\begin{table*}[tbp]
\caption{
Adopted analytical pairwise functions for $f_n (R)$ and their derivatives $f_n' (R)$. 
The internal parameters $a$ and $b$ are always positive except for parameter $a$ in the STO and GTO functions.
For the Bessel and Neumann functions, $n$ is a positive integer.
}
\label{lasso2:fn_table}
\begin{ruledtabular}
\begin{tabular}{lcc}
& $f_n(R)$ & $f_n'(R)$ \\
\hline
Bessel & $J_n (R)$ & $- f_1(R)$ ($n = 0$), $\left[ f_{n-1} (R) - f_{n+1} \right] / 2$ ($n \geq 1$)\\
Neumann & $Y_n (R)$ & $- f_1(R) $ ($n = 0$), $\left[ f_{n-1} (R) - f_{n+1} \right] / 2$ ($n \geq 1$)\\
\hline
Cosine & $\cos (a R) $ & $-a \sin (a R)$ \\
Modified Morlet wavelet (MMW) & $\cos(a R) / \cosh (R) $ & $ - a \sin (aR) / \cosh(R) - f_n(R) \tanh(R) $ \\
\hline
Gaussian & $\exp \left[ -a (R - b) ^2\right]$ & $-2a (R-b) f_n(R)$  \\
Slater-type orbital (STO) & $R^a \exp \left( -b R \right)$  & $(a-b R) R^{a-1} \exp \left( -b R \right)$  \\
Gaussian-type orbital (GTO) & $R^a \exp \left( -b R^2 \right)$  & $(a-2b R^2) R^{a-1} \exp \left( -b R^2 \right)$ \\
\end{tabular}
\end{ruledtabular}
\end{table*}

For the pairwise analytical function $f_n$, we introduce Gaussian, cosine, Bessel, Neumann, modified Morlet wavelet (MMW), Slater-type orbital (STO) and Gaussian-type orbital (GTO) functions.
Table \ref{lasso2:fn_table} shows the different function forms of $f_n$ and their derivatives with respect to the distance $f_n'$ used in this study.
The derivatives can be seen in the expressions for the forces acting on atoms and the stress tensor.
For the cosine and MMW types, function forms with a single internal parameter are introduced, while for the Gaussian, STO and GTO types, function forms with two internal parameters are used.
Using a number of functions with different internal parameters for each type of function, a systematic set of basis functions used to select the basis in elastic net regression is obtained.
For the cutoff function, we adopt the cosine-based cutoff function used in Ref. \onlinecite{behler2007generalized}, expressed as
\begin{eqnarray}
f_c(R) = \left\{
\begin{aligned}
& \frac{1}{2} \left[ \cos \left( \pi \frac{R}{R_c} \right) + 1\right] & (R \le R_c)\\
& 0 & (R > R_c)
\end{aligned}
\right . .
\end{eqnarray}

\section{Optimization of input factors}
\label{lasso2:preparation_input_data_section}
The accuracy of the elastic net PES mainly depends on 
1) the cutoff radius $R_c$, 
2) the size of the training data,
3) the variety of structures included in the training data,
4) the observation properties used for regression,
5) the candidate basis functions
and 
6) the parameters $\alpha$ and $\lambda$ in the minimization function of the elastic net regression.
However, it is difficult to optimize all of these input factors simultaneously.
We therefore optimize them in a stepwise manner.

\subsection{DFT data set}
To begin with, training and test data sets were generated from systematic DFT calculations.
The test data set was used to examine the predictive power for structures that were not included in the training data set.
We generated 2700 and 300 configurations for the training and test data sets, respectively, for each elemental metal.
They include structures made by isotropic expansions, random expansions and random distortions of ideal face-centered-cubic (fcc), body-centered-cubic (bcc), hexagonal-closed-packed (hcp), simple cubic (sc), $\omega$ and $\beta$-tin structures, in which the atomic positions and lattice constants were fully optimized.
Random structures were generated using Gaussian random numbers with ten different values for the variance.
These configurations were made using supercells constructed by the $2\times2\times2$, $3\times3\times3$, $3\times3\times3$, $4\times4\times4$, $3\times3\times3$ and $2\times2\times2$ expansions of the conventional unit cells for fcc, bcc, hcp, sc, $\omega$ and $\beta$-tin structures, which are composed of 32, 54, 54, 64, 81 and 32 atoms, respectively.

For a total of 3000 configurations for each elemental metal, DFT calculations were performed using the plane-wave basis projector augmented wave (PAW) method\cite{PAW1,PAW2} within the Perdew-Burke-Ernzerhof exchange-correlation functional\cite{GGA:PBE96} as implemented in the \textsc{vasp} code\cite{VASP1,VASP2}.
The cutoff energy was set to 400 eV.
The total energies converged to less than 10$^{-3}$ meV/supercell.
For only the ideal structures, the atomic positions and lattice constants were optimized until the residual forces became less than $10^{-3}$ eV/\AA.

\begin{table}[tbp]
\caption{
Optimal cutoff radius $R_c$. 
PESs are constructed by linear ridge regression using a given basis set composed of 180 basis functions including 18 Bessel, 18 Neumann, 60 cosine and 84 Gaussian functions.
}
\label{lasso2:cutoff_radii}
\begin{ruledtabular}
\begin{tabular}{ccccc}
& & \multicolumn{3}{c}{Linear ridge regression ($N_{\rm basis} = 180$)} \\
& $R_c$ & RMSE (energy) & RMSE (force) & RMSE (stress) \\
& (\AA) & (meV/atom) & (eV/\AA) & (GPa) \\
\hline
Ag & 7.5   & 2.4 & 0.012 & 0.08  \\ 
Al & 8.0   & 4.0 & 0.020 & 0.15  \\
Au & 6.0   & 3.0 & 0.030 & 0.16  \\
Ca & 9.5   & 1.2 & 0.011 & 0.03  \\
Cu & 7.5   & 2.6 & 0.018 & 0.10  \\
Ga & 10.0  & 1.9 & 0.019 & 0.11  \\
In & 10.0  & 1.9 & 0.017 & 0.07  \\
K  & 10.0  & 0.5 & 0.001 & 0.00  \\
Li & 8.5   & 0.5 & 0.005 & 0.02  \\
Zn & 10.0  & 3.0 & 0.021 & 0.22  \\
\end{tabular}
\end{ruledtabular}
\end{table}

\subsection{Cutoff radius}
The optimal cutoff radius was determined for each elemental metal using linear ridge regression with a given set of 180 basis functions consisting of 18 Bessel, 18 Neumann, 60 cosine and 84 Gaussian functions.
We searched for a cutoff radius giving a low prediction error by constructing PESs using several cutoff radii ranging from 5 to 10 \AA\ with intervals of 0.5 \AA.
The ridge penalty term was set to $\lambda = 10^{-4}$.
To estimate the prediction error of each PES, we calculated the root-mean-square error (RMSE) between the observation property predicted by the DFT calculation and that predicted by the PES for the test data.
Then, the convergence of the RMSE was examined.
Here, the energy, the force and the stress tensor were used as the observation properties. 
In addition, the cutoff radius also plays an essential role in expressing the energy of structures with large volumes.
However, the contribution of such structures to the RMSE was minor in our data sets.
Therefore, we determined the cutoff radii using the convergence of the energy-volume curve in addition to the convergence of the RMSE.

Table \ref{lasso2:cutoff_radii} shows the optimized cutoff radii.
Hereafter, these optimized values will be used.
Table \ref{lasso2:cutoff_radii} also shows the RMSEs of the energy, force and stress tensor.
We obtained PESs with the RMSEs for the energy ranging from 0.5 to 4.0 meV/atom.
In other words, PESs with high accuracy were obtained even using a given set of basis functions and linear ridge regression.
The given set of basis functions may be acceptable for expressing the energetics of the ten different elemental metals.
However, this may not be the case with other systems. 

\subsection{Observation property}
The dependence of the PES accuracy on the observation property of the training data was examined.
We compared two training sets of observation properties used for regression.
One training set was composed only of the energy, and the other set is composed of the energy, force and stress tensor.
The comparison was carried out using the above 180 basis functions, the optimized cutoff radius and linear ridge regression.
When using only the energy as the observation property, the RMSE was small for the energy but large for the force and the stress tensor.
Including the force and the stress tensor to the observation properties resulted in improved prediction for the force and the stress tensor at the expense of the predictive power for the energy.
To ensure accuracy for both the force and the stress tensor, which is essential for calculations of phonon dispersions and structure optimization, we hereafter use the energy, force and stress tensor as the observation properties unless otherwise specified.

\begin{figure*}[tbp]
\begin{center}
\includegraphics[width=0.7\linewidth,clip]{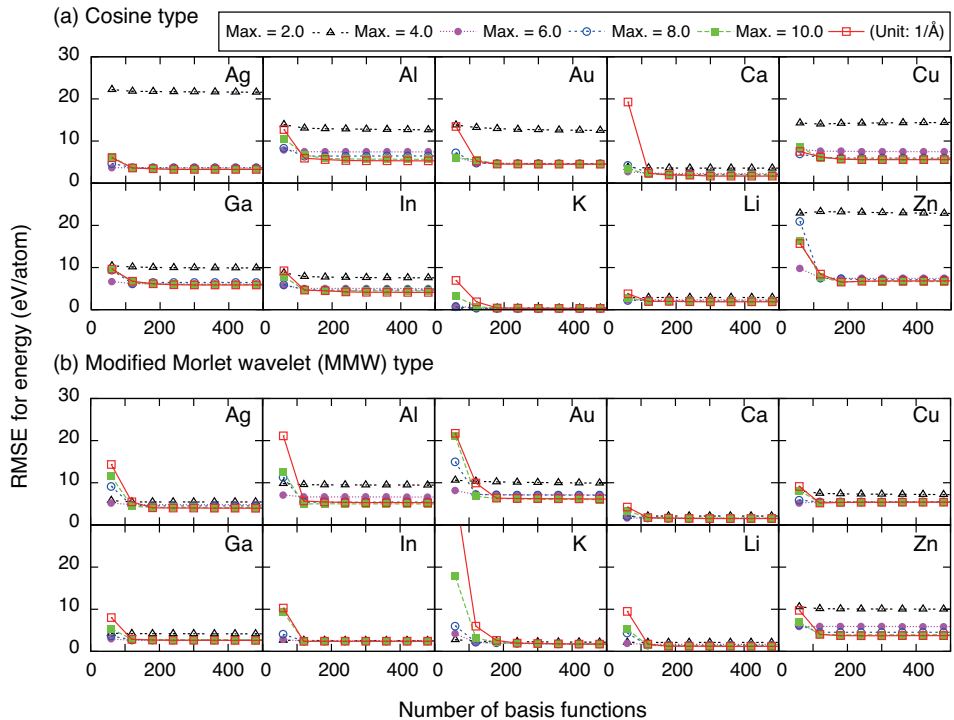} 
\caption{
Dependence of RMSE of linear ridge PES on the number of basis functions for (a) cosine and (b) MMW types.
Max. stands for the maximum value of the arithmetic sequence of the internal parameter.
}
\label{lasso2:candidate_basis_figure}
\end{center}
\end{figure*}

\begin{table}[tbp]
\caption{
Optimized candidate basis set used in elastic net regression.
Min. and Max. stand for the minimum and maximum values of the arithmetic sequence of the internal parameter, respectively.
$N_{\rm seq}$ denotes the number of components of the sequence.
}
\label{lasso2:candidate_basis_table}
\begin{ruledtabular}
\begin{tabular}{cccccc}
\multirow{2}{*}{Basis type} & Number of & \multicolumn{4}{c}{Internal parameter} \\
& basis functions & & Min. & Max. & $N_{\rm seq}$\\
\hline
Bessel & 18 & $n$ & 0 & 5 & 6 \\
Neumann & 18 & $n$ & 0 & 5 & 6 \\
\hline
Cosine & 300 & $a$ & 0.1 & 10.0 & 100 \\
MMW & 300 & $a$ & 0.1 & 10.0 & 100 \\
\hline
\multirow{2}{*}{Gaussian} & \multirow{2}{*}{1200} & $a$ & 0.1 & 2.0 & 20\\
& & $b$ & 0.0 & 5.0 & 20 \\
\multirow{2}{*}{STO} & \multirow{2}{*}{1500} & $a$ & -2 & 2 & 5 \\
& & $b$ & 0.1 & 10.0 & 100 \\
\multirow{2}{*}{GTO} & \multirow{2}{*}{1500} & $a$ & -2 & 2 & 5 \\
& & $b$ & 0.1 & 10.0 & 100\\
\hline
Total & 4836 & & & &\\
\end{tabular}
\end{ruledtabular}
\end{table}

\subsection{Candidate basis set}
Even using a combination of linear ridge regression and the above 180 basis functions, PESs with high accuracy were sometimes obtained.
However, a PES with a smaller number of basis functions is generally preferable to accelerate the computation of the energy, forces and stress tensors.
Additionally, it may be possible to find more suitable basis functions by considering other basis functions.
Therefore, it is useful to select suitable basis functions from a candidate basis set including other basis functions by elastic net regression.

In general, a candidate basis set should ideally be compact for the following reasons.
1) If the elastic net uses too many basis functions compared to the number of input observations, the selection of a good set of basis functions tends to be difficult.
2) The amount of available memory on computers can be exhausted, particularly when the forces and stress tensor are used as the observations.

A compact candidate basis set was obtained by optimization on the basis of the results of linear ridge regression with a single type of basis function.
For a basis type with a single internal parameter, a trial set for the internal parameter was given by an arithmetic sequence.
The sequence can be specified by the minimum and maximum values and the number of components of the sequence.
For the Bessel and Neumann types, we set the minimum and interval of the arithmetic sequence to zero and one, respectively. 
For the cosine and MMW functions, the minimum value was taken to be 0.1. 
Linear ridge PESs were then constructed using many trial sets for each basis type.
Here, all the basis functions with $p=1,2,3$ for each $f_n$ were considered because it has been shown that the use of $p=1,2,3$ terms greatly decreases the prediction error for elemental Na and Mg\cite{PhysRevB.90.024101}.
The ridge penalty term was taken to be $\lambda = 10^{-4}$.

Figures \ref{lasso2:candidate_basis_figure} (a) and (b) show the convergence of the RMSE for the energy with respect to the number of basis functions for the cosine and MMW types.
Unfixed parameters of the sequence were determined from the convergence of the RMSE.
Table \ref{lasso2:candidate_basis_table} shows the optimized number of basis functions together with the minimum and maximum values of the sequence. 
We also tested polynomial forms, three types of exponential forms and Mexican hat wavelets with a single internal parameter as candidates of $f_n$.
However, they showed a larger RMSE or unstable behavior when performing regression.

For each basis type with two internal parameters, two arithmetic sequences were given and all combinations of their components were considered.
For the Gaussian type, the minimum values of the sequences for $a$ and $b$ were fixed to 0.1 and 0.0, respectively. 
The number of components of each sequence was set to 20.
Therefore, each sequence was specified only by the maximum value.
For the STO and GTO types, the sequence for $a$ was fixed so that the minimum value, the interval and the maximum value were given as $-$2, 1 and 2, respectively.
The minimum value of the sequence for $b$ was set to 0.1.
The sequence for $b$ was specified by the maximum value and the number of components of the sequence.
Table \ref{lasso2:candidate_basis_table} shows the optimized sequences.
We used all the types of basis functions shown in Table \ref{lasso2:candidate_basis_table} as the candidate basis set.
The total number of functions in the candidate basis set was 4836. 

\begin{figure*}[tbp]
\begin{center}
\includegraphics[width=0.8\linewidth,clip]{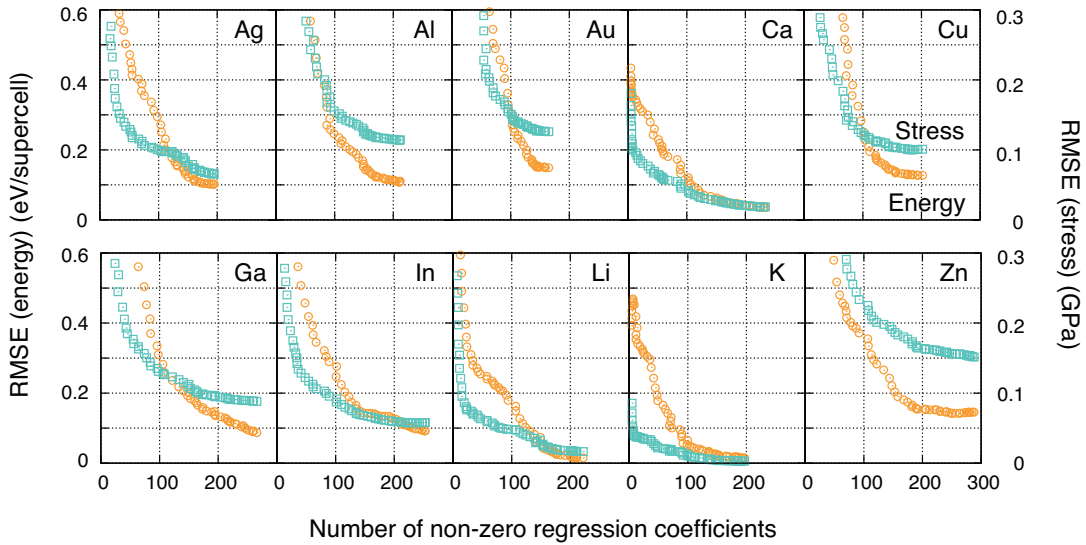} 
\caption{
Dependence of RMSEs for the energy and the stress tensor of elastic net PES ($\alpha = 1$, LASSO) on the number of the basis functions for ten elemental metals.
RMSEs for the energy and the stress tensor are shown by orange open circles and blue open squares, respectively.
}
\label{lasso2:elastic_net_rmse_figure}
\end{center}
\end{figure*}

\begin{table*}[tbp]
\caption{
RMSEs for the energy, the force and the stress tensor of elastic net PESs showing the minimum criterion score.
Equilibrium lattice constants for the bcc and fcc structures estimated from the elastic net PES are also shown. 
Values in brackets were obtained directly by DFT calculation.
}
\label{lasso2:elastic_net_rmse}
\begin{ruledtabular}
\begin{tabular}{ccccccc}
\multirow{2}{*}{Element} & Number of  & RMSE (energy) & RMSE (force) & RMSE (stress) & $a$ (bcc) & $a$ (fcc) \\
& basis functions & (meV/atom) & (eV/\AA) & (GPa) & (\AA) & (\AA) \\
\hline
Ag & 190 & 2.2 & 0.011 & 0.07 & 3.309 (3.311) & 4.157 (4.160) \\
Al & 210 & 3.5 & 0.020 & 0.12 & 3.234 (3.233) & 4.039 (4.038) \\
Au & 165 & 2.4 & 0.030 & 0.15 & 3.316 (3.309) & 4.172 (4.164) \\
Ca & 234 & 1.2 & 0.010 & 0.03 & 4.383 (4.381) & 5.522 (5.519) \\
Cu & 202 & 2.6 & 0.018 & 0.12 & 2.885 (2.887) & 3.630 (3.633) \\
Ga & 266 & 2.2 & 0.017 & 0.09 & 3.371 (3.371) & 4.227 (4.228) \\
In & 253 & 2.3 & 0.019 & 0.07 & 3.814 (3.815) & 4.797 (4.797) \\
K  & 197 & 0.3 & 0.001 & 0.00 & 5.284 (5.283) & 6.666 (6.662) \\
Li & 222 & 0.4 & 0.005 & 0.02 & 3.440 (3.439) & 4.329 (4.331) \\
Zn & 288 & 2.9 & 0.016 & 0.15 & 3.130 (3.136) & 3.928 (3.935) \\
\end{tabular}
\end{ruledtabular}
\end{table*}

\section{Elastic net potential energy surface}
\label{lasso2:elastic_net_section}
PESs were then estimated by elastic net regression using the DFT observations, the candidate basis set and the optimized cutoff radius.
PESs were optimized by changing the parameters $\alpha$ and $\lambda$ in the minimization function of Eqn. (\ref{lasso2:elastic_net_min_equation}).
We varied $\lambda$ from $10^3$ to $10^{-3}$ and adopted values of $\alpha$ of 0.6, 0.8 and 1.
In elastic net regression, we used only the energy and stress tensor as the observations because the available computational memory was limited.
Although the regression coefficients obtained by the elastic net were valid as they were, we reestimated them using linear ridge regression, where the energy, forces, and stress tensor were used as the observations.

As a criterion score to determine the optimal PES, the average of the RMSEs for the energy and the stress tensor was used.
Here, we regarded the PES with the lowest criterion score as the optimal one. 
Note, however, that the definition of the optimal PES depends on the purpose.
In another situation, a PES with a smaller number of basis functions may be regarded as the optimal one when a decrease in the computational costs at the expense of slight degradation of the accuracy is desired.

Figure \ref{lasso2:elastic_net_rmse_figure} shows the dependence of the RMSE for the energy and stress tensor on the number of basis functions when $\alpha = 1$.
The number of selected basis functions tended to increase with decreasing $\lambda$.
At the same time, the RMSE for the energy and stress tensor tended to decrease.
Although multiple PESs with the same number of basis functions were sometimes obtained from different values of $\lambda$, only the PES with the lowest criterion score among the PESs with the same number of basis functions is shown in Fig. \ref{lasso2:elastic_net_rmse_figure}.
On the other hand, the criterion score does not change significantly with decreasing $\alpha$ although the number of selected basis functions increases.
Therefore, we will hereafter show only the results for $\alpha=1$.

Table \ref{lasso2:elastic_net_rmse} shows the RMSEs for the energy, the force and the stress tensor of the optimal elastic net PES.
We obtained PESs with the RMSE for the energy in the range of 0.3$-$3.5 meV/atom for the ten elemental metals using only 165$-$288 basis functions.
The RMSEs for the force and the stress are within 0.03 eV/\AA\ and 0.15 GPa, respectively.
Compared with the RMSEs of the PESs constructed from the given basis set shown in Table \ref{lasso2:cutoff_radii}, the prediction errors are reduced for some elemental metals as a consequence of the automatic optimization of the basis set.
Table \ref{lasso2:elastic_net_rmse} also shows the equilibrium lattice constants of the bcc and fcc structures for the ten elemental metals predicted by the elastic net PES, together with those predicted by DFT.
The PESs have equilibrium lattice constants that are in agreement with those obtained by DFT.

\begin{figure}[tbp]
\begin{center}
\includegraphics[width=\linewidth,clip]{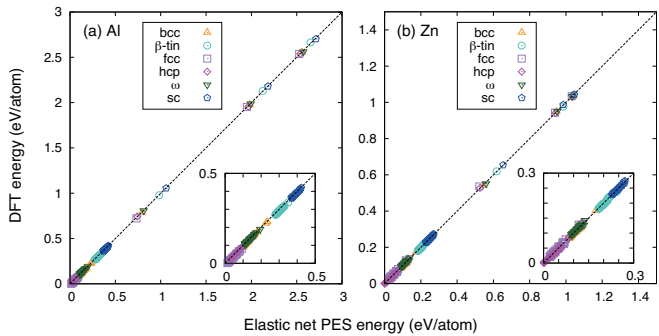} 
\caption{
Comparison of the energies predicted by the elastic net PES and DFT for (a) Al and (b) Zn, measured from the energy of the most stable structure among the bcc, fcc, hcp, sc, $\omega$ and $\beta$-tin structures.
}
\label{lasso2:error-figure}
\end{center}
\end{figure}

Figure \ref{lasso2:error-figure} shows a comparison of the energies of test data predicted by the elastic net PES and DFT for Al and Zn, showing the largest and second largest RMSEs for the energy.
As can be seen in Fig. \ref{lasso2:error-figure}, there is little difference between the DFT and elastic net PES energies regardless of the crystal structure.
In addition, no dependence of the RMSE on the energy can be clearly observed despite the wide range of structures included in both the training and test data.

\begin{figure}[tbp]
\begin{center}
\includegraphics[width=1.0\linewidth,clip]{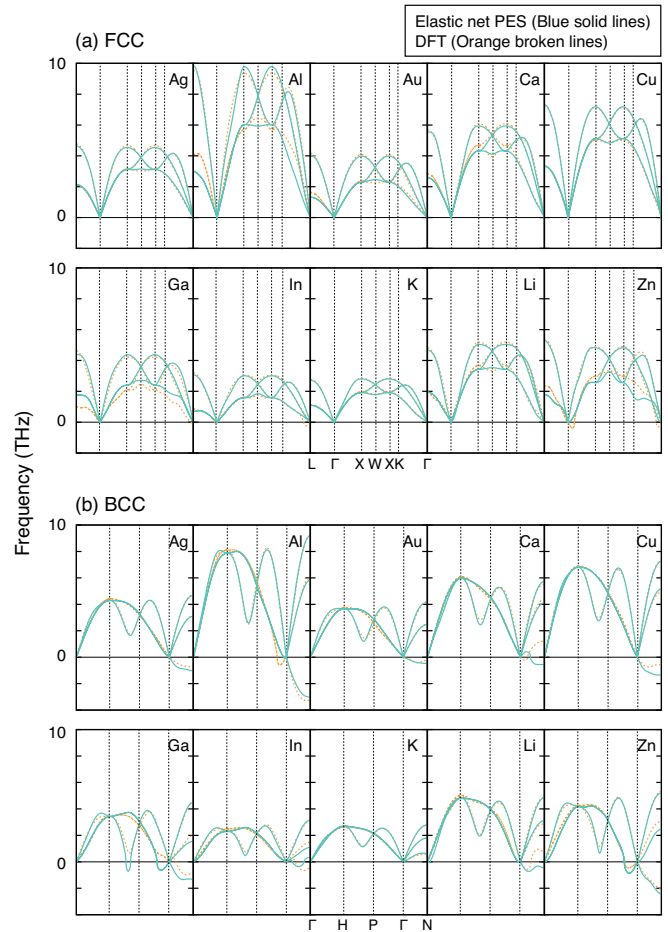} 
\caption{
Phonon dispersion relationships for ten elemental metals with (a) fcc and (b) bcc structures.
Phonon dispersion curves obtained by the elastic net PES and DFT are shown by blue solid and orange broken lines, respectively.
Negative values indicate imaginary modes.
}
\label{lasso2:phonon}
\end{center}
\end{figure}

The applicability of the elastic net PES to the calculation of the force was also examined by comparing phonon dispersion relationships computed by the elastic net PES and DFT.
The phonon dispersion relationships were calculated by the supercell approach\cite{parlinski1997first} for the bcc and fcc structures with the equilibrium lattice constant.
To evaluate the dynamical matrix, each symmetrically independent atomic position was displaced by 0.01 \AA.
The forces acting on atoms by the elastic net PES can then be analytically computed using Eqn. (\ref{lasso2:force-linear-model}).
Supercells were made by $4\times4\times4$ expansion of the conventional unit cells for both the bcc and fcc structures.
The phonon calculations were performed using the \textsc{phonopy} code\cite{PhysRevB.78.134106}.
Figure \ref{lasso2:phonon} shows the phonon dispersion relationships of the (a) bcc and (b) fcc structures for the ten elemental metals, computed by both the elastic net PES and DFT.
For all elemental metals with both the bcc and fcc structures, the phonon dispersion relationships calculated by the elastic net PES are in good agreement with those calculated by DFT.
This demonstrates that the elastic net PES is sufficiently accurate to perform atomistic simulations with similar accuracy to DFT calculation.

\section{Conclusion}
\label{lasso2:conclusion_section}
We have applied a method of constructing a linearized PES by elastic net regression to a wide range of elemental metals.
Compared with the other approach based on systematic first-principles calculations, the elastic net interatomic potential has the following main advantages.
1) A well-optimized sparse representation for the PES can be obtained, which increases the accuracy of atomistic simulations while decreasing the computational cost.
2) The accuracy can be easily controlled, i.e., the trade-off between the accuracy and computational cost is determined by a small number of parameters.
3) Information on the forces acting on atoms and stress tensors can be included in the training data in a straightforward manner.
This ensures the reliability of the force and stress tensor calculation using constructed interatomic potentials.

As a result of applying the present method, we found that the energetics can be expressed by a linear relationship with simple basis functions depending only on distances between atoms. 
A sparse set of suitable basis functions for expressing the PES can also be easily extracted from 4836 basis functions by elastic net regression.
As a result, we have obtained a sparse PES with prediction errors ranging from 0.3 to 3.5 meV/atom.
The prediction errors for the force and the stress tensor were within 0.03 eV/\AA\ and 0.15 GPa, respectively.
Also, we compared equilibrium lattice constants and phonon dispersion relationships obtained by the elastic net PES and by DFT calculation.
The former were in good agreement with the latter for all ten elemental metals considering in this study.

\begin{acknowledgments}
This study was supported by a Grant-in-Aid for Scientific Research (A) and a Grant-in-Aid for Scientific Research on Innovative Areas ``Nano Informatics'' (grant number 25106005) from Japan Society for the Promotion of Science (JSPS).
AS also acknowledges a Grant-in-Aid for Scientific Research (B) from JSPS.
\end{acknowledgments}

\appendix

\section{Expressions for forces acting on atoms and stress tensor}
\label{lasso2:force-stress-section}

Here, expressions for forces acting on atoms and the stress tensor are derived from the derivative of Eqn. (\ref{lasso2:energy_equation}) with respect to the atomic positions provided in Cartesian coordinates.
Although they depend on the form of the basis functions, expressions for a linear model involving only basis functions depending only on pair distances are derived.
Since the total energy of structure $i$ has a linear relationship with the sum of the basis functions,
$\alpha$th component of the force acting on atom $l$ of structure $i$ is expressed by
\begin{eqnarray}
F^{(i)}_{l,\alpha} & = &- \frac{\partial E^{(i)}}{\partial R_{l,\alpha}^{(i)}} \nonumber \\
& = & - \sum_{n,p} w_{n,p} \frac{\partial x^{(i)}_{n,p}}{\partial  R_{l,\alpha}^{(i)}} \nonumber \\
& = & \sum_{n,p} w_{n,p} x_{{\rm force},n,p}^{(i,l,\alpha)} ,
\label{lasso2:force_def}
\end{eqnarray}
where
\begin{eqnarray}
x_{{\rm force},n,p}^{(i,l,\alpha)} & = & - \frac{\partial x^{(i)}_{n,p}}{\partial  R_{l,\alpha}^{(i)}} \nonumber \\
& = & - p \sum_j b^{(i,j)}_{n,p-1} \frac{\partial b^{(i,j)}_{n,1}}{\partial  R_{l,\alpha}^{(i)}}.
\label{lasso2:basis_force_def}
\end{eqnarray}
The stress tensor is generally obtained by virial stress computation.
The virial stress tensor $\sigma_{\alpha\beta}$ is expressed as
\begin{equation}
\sigma_{\alpha\beta}^{(i)} = \frac{1}{V} \sum_l R_{l,\alpha}^{(i)} F_{l,\beta}^{(i)},
\end{equation}
where $V$ denotes the volume of the cell containing $N^{(i)}$ atoms.
Using the expression for the forces in Eqn. (\ref{lasso2:force_def}), the stress tensor is derived as the following linear equation:
\begin{eqnarray}
\sigma_{\alpha\beta}^{(i)} & = & - \frac{1}{V} \sum_l R_{l,\alpha}^{(i)} \left[ \sum_{n,p} w_{n,p} \left( p \sum_j b^{(i,j)}_{n,p-1} \frac{\partial b^{(i,j)}_{n,1}}{\partial  R_{l,\beta}^{(i)}} \right) \right] \nonumber \\
& = & \sum_{n,p} w_{n,p} \left[ - \frac{p}{V} \sum_l R_{l,\alpha}^{(i)} \sum_j b^{(i,j)}_{n,p-1} \frac{\partial b^{(i,j)}_{n,1}}{\partial  R_{l,\beta}^{(i)}} \right] \nonumber \\
& = & \sum_{n,p} w_{n,p} x_{{\rm stress},n,p}^{(i,\alpha,\beta)},
\label{lasso2:stress_eq}
\end{eqnarray}
where
\begin{equation}
x_{{\rm stress},n,p}^{(i,\alpha,\beta)} = - \frac{p}{V} \sum_l R_{l,\alpha}^{(i)} \sum_j b^{(i,j)}_{n,p-1} \frac{\partial b^{(i,j)}_{n,1}}{\partial  R_{l,\beta}^{(i)}}.
\end{equation}
By computing all the contributions from atoms within the cutoff radius using Eqn. (\ref{lasso2:stress_eq}), the virial stress is obtained.

The derivative of the basis functions with respect to the $\alpha$th component of the atomic position is written as
\begin{widetext}
\begin{eqnarray}
\frac{\partial b^{(j)}_{n,1}}{\partial  R_{l,\alpha}} & = & \sum_k \left[ \frac{\partial f_n (R_{jk})}{\partial  R_{l,\alpha}} f_c(R_{jk}) + f_n (R_{jk}) \frac{\partial f_c (R_{jk})}{\partial  R_{l,\alpha}} \right] \nonumber \\
& = & \sum_k \left[f_n'(R_{jk}) f_c(R_{jk}) + f_n (R_{jk}) f_c' (R_{jk}) \right] \frac{\partial R_{jk}}{\partial  R_{l,\alpha}}, 
\label{lasso2:basis_diff}
\end{eqnarray}
\end{widetext}
where structure index $i$ is omitted.
Three types of derivatives can be found in Eqn. (\ref{lasso2:basis_diff}).
The derivative of the distance with respect to the $\alpha$th component of the atomic position is expressed as
\begin{eqnarray}
\frac{\partial R_{jk}}{\partial R_{l,\alpha}} =  \left\{
\begin{aligned}
&\frac{\left(\bm{R}_j - \bm{R}_k\right)_\alpha}{R_{jk}} \:\:\: (l = j) \\
&\frac{\left(\bm{R}_k - \bm{R}_j\right)_\alpha}{R_{jk}} \:\:\: (l = k)
\end{aligned}
\right . ,
\end{eqnarray}
where $\bm{R}_j$ denotes the three-dimensional atomic position of atom $j$ in Cartesian coordinates.
The derivative of the cutoff function with respect to the distance is given by 
\begin{equation}
f_c' (R_{jk}) = -\frac{\pi}{2 R_c} \sin \left(\pi \frac{R_{jk}}{R_c} \right).
\end{equation}
The derivative of the pairwise function $f_n$ with respect to the distance depends on the selection of the functions, hence the expression for the derivative for each type of $f_n$ is shown in Table \ref{lasso2:fn_table}.

\bibliography{lasso2}

\begin{thebibliography}{57}%
\makeatletter
\providecommand \@ifxundefined [1]{%
 \@ifx{#1\undefined}
}%
\providecommand \@ifnum [1]{%
 \ifnum #1\expandafter \@firstoftwo
 \else \expandafter \@secondoftwo
 \fi
}%
\providecommand \@ifx [1]{%
 \ifx #1\expandafter \@firstoftwo
 \else \expandafter \@secondoftwo
 \fi
}%
\providecommand \natexlab [1]{#1}%
\providecommand \enquote  [1]{``#1''}%
\providecommand \bibnamefont  [1]{#1}%
\providecommand \bibfnamefont [1]{#1}%
\providecommand \citenamefont [1]{#1}%
\providecommand \href@noop [0]{\@secondoftwo}%
\providecommand \href [0]{\begingroup \@sanitize@url \@href}%
\providecommand \@href[1]{\@@startlink{#1}\@@href}%
\providecommand \@@href[1]{\endgroup#1\@@endlink}%
\providecommand \@sanitize@url [0]{\catcode `\\12\catcode `\$12\catcode
  `\&12\catcode `\#12\catcode `\^12\catcode `\_12\catcode `\%12\relax}%
\providecommand \@@startlink[1]{}%
\providecommand \@@endlink[0]{}%
\providecommand \url  [0]{\begingroup\@sanitize@url \@url }%
\providecommand \@url [1]{\endgroup\@href {#1}{\urlprefix }}%
\providecommand \urlprefix  [0]{URL }%
\providecommand \Eprint [0]{\href }%
\providecommand \doibase [0]{http://dx.doi.org/}%
\providecommand \selectlanguage [0]{\@gobble}%
\providecommand \bibinfo  [0]{\@secondoftwo}%
\providecommand \bibfield  [0]{\@secondoftwo}%
\providecommand \translation [1]{[#1]}%
\providecommand \BibitemOpen [0]{}%
\providecommand \bibitemStop [0]{}%
\providecommand \bibitemNoStop [0]{.\EOS\space}%
\providecommand \EOS [0]{\spacefactor3000\relax}%
\providecommand \BibitemShut  [1]{\csname bibitem#1\endcsname}%
\let\auto@bib@innerbib\@empty
\bibitem [{\citenamefont {Alder}\ and\ \citenamefont
  {Wainwright}(1959)}]{alder1959studies}%
  \BibitemOpen
  \bibfield  {author} {\bibinfo {author} {\bibfnamefont {B.~J.}\ \bibnamefont
  {Alder}}\ and\ \bibinfo {author} {\bibfnamefont {T.~E.}\ \bibnamefont
  {Wainwright}},\ }\href@noop {} {\bibfield  {journal} {\bibinfo  {journal} {J.
  Chem. Phys.}\ }\textbf {\bibinfo {volume} {31}},\ \bibinfo {pages} {459}
  (\bibinfo {year} {1959})}\BibitemShut {NoStop}%
\bibitem [{\citenamefont {Car}\ and\ \citenamefont
  {Parrinello}(1985)}]{car1985unified}%
  \BibitemOpen
  \bibfield  {author} {\bibinfo {author} {\bibfnamefont {R.}~\bibnamefont
  {Car}}\ and\ \bibinfo {author} {\bibfnamefont {M.}~\bibnamefont
  {Parrinello}},\ }\href@noop {} {\bibfield  {journal} {\bibinfo  {journal}
  {Phys. Rev. Lett.}\ }\textbf {\bibinfo {volume} {55}},\ \bibinfo {pages}
  {2471} (\bibinfo {year} {1985})}\BibitemShut {NoStop}%
\bibitem [{\citenamefont {Behler}\ and\ \citenamefont
  {Parrinello}(2007)}]{behler2007generalized}%
  \BibitemOpen
  \bibfield  {author} {\bibinfo {author} {\bibfnamefont {J.}~\bibnamefont
  {Behler}}\ and\ \bibinfo {author} {\bibfnamefont {M.}~\bibnamefont
  {Parrinello}},\ }\href@noop {} {\bibfield  {journal} {\bibinfo  {journal}
  {Phys. Rev. Lett.}\ }\textbf {\bibinfo {volume} {98}},\ \bibinfo {pages}
  {146401} (\bibinfo {year} {2007})}\BibitemShut {NoStop}%
\bibitem [{\citenamefont {Bart{\'o}k}\ \emph {et~al.}(2010)\citenamefont
  {Bart{\'o}k}, \citenamefont {Payne}, \citenamefont {Kondor},\ and\
  \citenamefont {Cs{\'a}nyi}}]{bartok2010gaussian}%
  \BibitemOpen
  \bibfield  {author} {\bibinfo {author} {\bibfnamefont {A.~P.}\ \bibnamefont
  {Bart{\'o}k}}, \bibinfo {author} {\bibfnamefont {M.~C.}\ \bibnamefont
  {Payne}}, \bibinfo {author} {\bibfnamefont {R.}~\bibnamefont {Kondor}}, \
  and\ \bibinfo {author} {\bibfnamefont {G.}~\bibnamefont {Cs{\'a}nyi}},\
  }\href@noop {} {\bibfield  {journal} {\bibinfo  {journal} {Phys. Rev. Lett.}\
  }\textbf {\bibinfo {volume} {104}},\ \bibinfo {pages} {136403} (\bibinfo
  {year} {2010})}\BibitemShut {NoStop}%
\bibitem [{\citenamefont {Bowman}\ \emph {et~al.}(1986)\citenamefont {Bowman},
  \citenamefont {Bittman},\ and\ \citenamefont {Harding}}]{Bowman1986}%
  \BibitemOpen
  \bibfield  {author} {\bibinfo {author} {\bibfnamefont {J.~M.}\ \bibnamefont
  {Bowman}}, \bibinfo {author} {\bibfnamefont {J.~S.}\ \bibnamefont {Bittman}},
  \ and\ \bibinfo {author} {\bibfnamefont {L.~B.}\ \bibnamefont {Harding}},\
  }\href {\doibase http://dx.doi.org/10.1063/1.451246} {\bibfield  {journal}
  {\bibinfo  {journal} {J. Chem. Phys.}\ }\textbf {\bibinfo {volume} {85}},\
  \bibinfo {pages} {911} (\bibinfo {year} {1986})}\BibitemShut {NoStop}%
\bibitem [{\citenamefont {Chapman}\ \emph {et~al.}(1983)\citenamefont
  {Chapman}, \citenamefont {Dupuis},\ and\ \citenamefont
  {Green}}]{Chapman198393}%
  \BibitemOpen
  \bibfield  {author} {\bibinfo {author} {\bibfnamefont {S.}~\bibnamefont
  {Chapman}}, \bibinfo {author} {\bibfnamefont {M.}~\bibnamefont {Dupuis}}, \
  and\ \bibinfo {author} {\bibfnamefont {S.}~\bibnamefont {Green}},\ }\href
  {\doibase http://dx.doi.org/10.1016/0301-0104(83)87010-4} {\bibfield
  {journal} {\bibinfo  {journal} {Chem. Phys.}\ }\textbf {\bibinfo {volume}
  {78}},\ \bibinfo {pages} {93} (\bibinfo {year} {1983})}\BibitemShut {NoStop}%
\bibitem [{\citenamefont {Dawes}\ \emph {et~al.}(2007)\citenamefont {Dawes},
  \citenamefont {Thompson}, \citenamefont {Guo}, \citenamefont {Wagner},\ and\
  \citenamefont {Minkoff}}]{Dawes2007_1.2730798}%
  \BibitemOpen
  \bibfield  {author} {\bibinfo {author} {\bibfnamefont {R.}~\bibnamefont
  {Dawes}}, \bibinfo {author} {\bibfnamefont {D.~L.}\ \bibnamefont {Thompson}},
  \bibinfo {author} {\bibfnamefont {Y.}~\bibnamefont {Guo}}, \bibinfo {author}
  {\bibfnamefont {A.~F.}\ \bibnamefont {Wagner}}, \ and\ \bibinfo {author}
  {\bibfnamefont {M.}~\bibnamefont {Minkoff}},\ }\href {\doibase
  http://dx.doi.org/10.1063/1.2730798} {\bibfield  {journal} {\bibinfo
  {journal} {J. Chem. Phys.}\ }\textbf {\bibinfo {volume} {126}},\ \bibinfo
  {eid} {184108} (\bibinfo {year} {2007})}\BibitemShut {NoStop}%
\bibitem [{\citenamefont {Dawes}\ \emph {et~al.}(2008)\citenamefont {Dawes},
  \citenamefont {Thompson}, \citenamefont {Wagner},\ and\ \citenamefont
  {Minkoff}}]{Dowes2008_1.2831790}%
  \BibitemOpen
  \bibfield  {author} {\bibinfo {author} {\bibfnamefont {R.}~\bibnamefont
  {Dawes}}, \bibinfo {author} {\bibfnamefont {D.~L.}\ \bibnamefont {Thompson}},
  \bibinfo {author} {\bibfnamefont {A.~F.}\ \bibnamefont {Wagner}}, \ and\
  \bibinfo {author} {\bibfnamefont {M.}~\bibnamefont {Minkoff}},\ }\href
  {\doibase http://dx.doi.org/10.1063/1.2831790} {\bibfield  {journal}
  {\bibinfo  {journal} {J. Chem. Phys.}\ }\textbf {\bibinfo {volume} {128}},\
  \bibinfo {eid} {084107} (\bibinfo {year} {2008})}\BibitemShut {NoStop}%
\bibitem [{\citenamefont {Ischtwan}\ and\ \citenamefont
  {Collins}(1994)}]{Ischtwan1994}%
  \BibitemOpen
  \bibfield  {author} {\bibinfo {author} {\bibfnamefont {J.}~\bibnamefont
  {Ischtwan}}\ and\ \bibinfo {author} {\bibfnamefont {M.~A.}\ \bibnamefont
  {Collins}},\ }\href {\doibase http://dx.doi.org/10.1063/1.466801} {\bibfield
  {journal} {\bibinfo  {journal} {J. Chem. Phys.}\ }\textbf {\bibinfo {volume}
  {100}},\ \bibinfo {pages} {8080} (\bibinfo {year} {1994})}\BibitemShut
  {NoStop}%
\bibitem [{\citenamefont {Jordan}\ \emph {et~al.}(1995)\citenamefont {Jordan},
  \citenamefont {Thompson},\ and\ \citenamefont
  {Collins}}]{Jordan1995_1.469296}%
  \BibitemOpen
  \bibfield  {author} {\bibinfo {author} {\bibfnamefont {M.~J.~T.}\
  \bibnamefont {Jordan}}, \bibinfo {author} {\bibfnamefont {K.~C.}\
  \bibnamefont {Thompson}}, \ and\ \bibinfo {author} {\bibfnamefont {M.~A.}\
  \bibnamefont {Collins}},\ }\href {\doibase
  http://dx.doi.org/10.1063/1.469296} {\bibfield  {journal} {\bibinfo
  {journal} {J. Chem. Phys.}\ }\textbf {\bibinfo {volume} {102}},\ \bibinfo
  {pages} {5647} (\bibinfo {year} {1995})}\BibitemShut {NoStop}%
\bibitem [{\citenamefont {Fournier}\ and\ \citenamefont
  {Orel}(2013)}]{Fournier2013_1.4846297}%
  \BibitemOpen
  \bibfield  {author} {\bibinfo {author} {\bibfnamefont {R.}~\bibnamefont
  {Fournier}}\ and\ \bibinfo {author} {\bibfnamefont {S.}~\bibnamefont
  {Orel}},\ }\href {\doibase http://dx.doi.org/10.1063/1.4846297} {\bibfield
  {journal} {\bibinfo  {journal} {J. Chem. Phys.}\ }\textbf {\bibinfo {volume}
  {139}},\ \bibinfo {eid} {234110} (\bibinfo {year} {2013})}\BibitemShut
  {NoStop}%
\bibitem [{\citenamefont {Lorenz}\ \emph {et~al.}(2004)\citenamefont {Lorenz},
  \citenamefont {Gro{\ss}},\ and\ \citenamefont {Scheffler}}]{Lorenz2004210}%
  \BibitemOpen
  \bibfield  {author} {\bibinfo {author} {\bibfnamefont {S.}~\bibnamefont
  {Lorenz}}, \bibinfo {author} {\bibfnamefont {A.}~\bibnamefont {Gro{\ss}}}, \
  and\ \bibinfo {author} {\bibfnamefont {M.}~\bibnamefont {Scheffler}},\ }\href
  {\doibase http://dx.doi.org/10.1016/j.cplett.2004.07.076} {\bibfield
  {journal} {\bibinfo  {journal} {Chem. Phys. Lett.}\ }\textbf {\bibinfo
  {volume} {395}},\ \bibinfo {pages} {210 } (\bibinfo {year}
  {2004})}\BibitemShut {NoStop}%
\bibitem [{\citenamefont {Gassner}\ \emph {et~al.}(1998)\citenamefont
  {Gassner}, \citenamefont {Probst}, \citenamefont {Lauenstein},\ and\
  \citenamefont {Hermansson}}]{Gassuner1998_jp972209d}%
  \BibitemOpen
  \bibfield  {author} {\bibinfo {author} {\bibfnamefont {H.}~\bibnamefont
  {Gassner}}, \bibinfo {author} {\bibfnamefont {M.}~\bibnamefont {Probst}},
  \bibinfo {author} {\bibfnamefont {A.}~\bibnamefont {Lauenstein}}, \ and\
  \bibinfo {author} {\bibfnamefont {K.}~\bibnamefont {Hermansson}},\ }\href
  {\doibase 10.1021/jp972209d} {\bibfield  {journal} {\bibinfo  {journal} {J.
  Phys. Chem. A}\ }\textbf {\bibinfo {volume} {102}},\ \bibinfo {pages} {4596}
  (\bibinfo {year} {1998})}\BibitemShut {NoStop}%
\bibitem [{\citenamefont {Prudente}\ \emph {et~al.}(1998)\citenamefont
  {Prudente}, \citenamefont {Acioli},\ and\ \citenamefont
  {Neto}}]{Prudente1998_1.477550}%
  \BibitemOpen
  \bibfield  {author} {\bibinfo {author} {\bibfnamefont {F.~V.}\ \bibnamefont
  {Prudente}}, \bibinfo {author} {\bibfnamefont {P.~H.}\ \bibnamefont
  {Acioli}}, \ and\ \bibinfo {author} {\bibfnamefont {J.~J.~S.}\ \bibnamefont
  {Neto}},\ }\href {\doibase http://dx.doi.org/10.1063/1.477550} {\bibfield
  {journal} {\bibinfo  {journal} {J. Chem. Phys.}\ }\textbf {\bibinfo {volume}
  {109}},\ \bibinfo {pages} {8801} (\bibinfo {year} {1998})}\BibitemShut
  {NoStop}%
\bibitem [{\citenamefont {Prudente}\ and\ \citenamefont
  {Neto}(1998)}]{Prudente1998585}%
  \BibitemOpen
  \bibfield  {author} {\bibinfo {author} {\bibfnamefont {F.~V.}\ \bibnamefont
  {Prudente}}\ and\ \bibinfo {author} {\bibfnamefont {J.~J.~S.}\ \bibnamefont
  {Neto}},\ }\href {\doibase http://dx.doi.org/10.1016/S0009-2614(98)00207-3}
  {\bibfield  {journal} {\bibinfo  {journal} {Chem. Phys. Lett.}\ }\textbf
  {\bibinfo {volume} {287}},\ \bibinfo {pages} {585} (\bibinfo {year}
  {1998})}\BibitemShut {NoStop}%
\bibitem [{\citenamefont {Brown}\ \emph {et~al.}(1996)\citenamefont {Brown},
  \citenamefont {Gibbs},\ and\ \citenamefont {Clary}}]{Brown1996_1.472596}%
  \BibitemOpen
  \bibfield  {author} {\bibinfo {author} {\bibfnamefont {D.~F.~R.}\
  \bibnamefont {Brown}}, \bibinfo {author} {\bibfnamefont {M.~N.}\ \bibnamefont
  {Gibbs}}, \ and\ \bibinfo {author} {\bibfnamefont {D.~C.}\ \bibnamefont
  {Clary}},\ }\href {\doibase http://dx.doi.org/10.1063/1.472596} {\bibfield
  {journal} {\bibinfo  {journal} {J. Chem. Phys.}\ }\textbf {\bibinfo {volume}
  {105}},\ \bibinfo {pages} {7597} (\bibinfo {year} {1996})}\BibitemShut
  {NoStop}%
\bibitem [{\citenamefont {Blank}\ \emph {et~al.}(1995)\citenamefont {Blank},
  \citenamefont {Brown}, \citenamefont {Calhoun},\ and\ \citenamefont
  {Doren}}]{Blank1995_1.469597}%
  \BibitemOpen
  \bibfield  {author} {\bibinfo {author} {\bibfnamefont {T.~B.}\ \bibnamefont
  {Blank}}, \bibinfo {author} {\bibfnamefont {S.~D.}\ \bibnamefont {Brown}},
  \bibinfo {author} {\bibfnamefont {A.~W.}\ \bibnamefont {Calhoun}}, \ and\
  \bibinfo {author} {\bibfnamefont {D.~J.}\ \bibnamefont {Doren}},\ }\href
  {\doibase http://dx.doi.org/10.1063/1.469597} {\bibfield  {journal} {\bibinfo
   {journal} {J. Chem. Phys.}\ }\textbf {\bibinfo {volume} {103}},\ \bibinfo
  {pages} {4129} (\bibinfo {year} {1995})}\BibitemShut {NoStop}%
\bibitem [{\citenamefont {Sumpter}\ and\ \citenamefont
  {Noid}(1992)}]{Sumpter1992455}%
  \BibitemOpen
  \bibfield  {author} {\bibinfo {author} {\bibfnamefont {B.~G.}\ \bibnamefont
  {Sumpter}}\ and\ \bibinfo {author} {\bibfnamefont {D.~W.}\ \bibnamefont
  {Noid}},\ }\href {\doibase http://dx.doi.org/10.1016/0009-2614(92)85498-Y}
  {\bibfield  {journal} {\bibinfo  {journal} {Chem. Phys. Lett.}\ }\textbf
  {\bibinfo {volume} {192}},\ \bibinfo {pages} {455} (\bibinfo {year}
  {1992})}\BibitemShut {NoStop}%
\bibitem [{\citenamefont {Behler}\ \emph {et~al.}(2007)\citenamefont {Behler},
  \citenamefont {Lorenz},\ and\ \citenamefont {Reuter}}]{Behler2007_1.2746232}%
  \BibitemOpen
  \bibfield  {author} {\bibinfo {author} {\bibfnamefont {J.}~\bibnamefont
  {Behler}}, \bibinfo {author} {\bibfnamefont {S.}~\bibnamefont {Lorenz}}, \
  and\ \bibinfo {author} {\bibfnamefont {K.}~\bibnamefont {Reuter}},\ }\href
  {\doibase http://dx.doi.org/10.1063/1.2746232} {\bibfield  {journal}
  {\bibinfo  {journal} {J. Chem. Phys.}\ }\textbf {\bibinfo {volume} {127}},\
  \bibinfo {eid} {014705} (\bibinfo {year} {2007})}\BibitemShut {NoStop}%
\bibitem [{\citenamefont {Li}\ \emph {et~al.}(2013)\citenamefont {Li},
  \citenamefont {Jiang},\ and\ \citenamefont {Guo}}]{Jun2014_1.4832697}%
  \BibitemOpen
  \bibfield  {author} {\bibinfo {author} {\bibfnamefont {J.}~\bibnamefont
  {Li}}, \bibinfo {author} {\bibfnamefont {B.}~\bibnamefont {Jiang}}, \ and\
  \bibinfo {author} {\bibfnamefont {H.}~\bibnamefont {Guo}},\ }\href {\doibase
  http://dx.doi.org/10.1063/1.4832697} {\bibfield  {journal} {\bibinfo
  {journal} {J. Chem. Phys.}\ }\textbf {\bibinfo {volume} {139}},\ \bibinfo
  {eid} {204103} (\bibinfo {year} {2013})}\BibitemShut {NoStop}%
\bibitem [{\citenamefont {Jiang}\ and\ \citenamefont
  {Guo}(2014)}]{Bin2014_1.4887363}%
  \BibitemOpen
  \bibfield  {author} {\bibinfo {author} {\bibfnamefont {B.}~\bibnamefont
  {Jiang}}\ and\ \bibinfo {author} {\bibfnamefont {H.}~\bibnamefont {Guo}},\
  }\href {\doibase http://dx.doi.org/10.1063/1.4887363} {\bibfield  {journal}
  {\bibinfo  {journal} {J. Chem. Phys.}\ }\textbf {\bibinfo {volume} {141}},\
  \bibinfo {eid} {034109} (\bibinfo {year} {2014})}\BibitemShut {NoStop}%
\bibitem [{\citenamefont {Handley}\ and\ \citenamefont
  {Popelier}(2010)}]{Handley2010_jp9105585}%
  \BibitemOpen
  \bibfield  {author} {\bibinfo {author} {\bibfnamefont {C.~M.}\ \bibnamefont
  {Handley}}\ and\ \bibinfo {author} {\bibfnamefont {P.~L.~A.}\ \bibnamefont
  {Popelier}},\ }\href {\doibase 10.1021/jp9105585} {\bibfield  {journal}
  {\bibinfo  {journal} {J. Phys. Chem. A}\ }\textbf {\bibinfo {volume} {114}},\
  \bibinfo {pages} {3371} (\bibinfo {year} {2010})}\BibitemShut {NoStop}%
\bibitem [{\citenamefont {Agrawal}\ \emph {et~al.}(2009)\citenamefont
  {Agrawal}, \citenamefont {Malshe}, \citenamefont {Narulkar}, \citenamefont
  {Raff}, \citenamefont {Hagan}, \citenamefont {Bukkapatnum},\ and\
  \citenamefont {Komanduri}}]{Agrawal2009_jp8085232}%
  \BibitemOpen
  \bibfield  {author} {\bibinfo {author} {\bibfnamefont {P.~M.}\ \bibnamefont
  {Agrawal}}, \bibinfo {author} {\bibfnamefont {M.}~\bibnamefont {Malshe}},
  \bibinfo {author} {\bibfnamefont {R.}~\bibnamefont {Narulkar}}, \bibinfo
  {author} {\bibfnamefont {L.~M.}\ \bibnamefont {Raff}}, \bibinfo {author}
  {\bibfnamefont {M.}~\bibnamefont {Hagan}}, \bibinfo {author} {\bibfnamefont
  {S.}~\bibnamefont {Bukkapatnum}}, \ and\ \bibinfo {author} {\bibfnamefont
  {R.}~\bibnamefont {Komanduri}},\ }\href {\doibase 10.1021/jp8085232}
  {\bibfield  {journal} {\bibinfo  {journal} {J. Phys. Chem. A}\ }\textbf
  {\bibinfo {volume} {113}},\ \bibinfo {pages} {869} (\bibinfo {year}
  {2009})}\BibitemShut {NoStop}%
\bibitem [{\citenamefont {Nguyen}\ and\ \citenamefont
  {Le}(2012)}]{Nguyen2012_jp3020386}%
  \BibitemOpen
  \bibfield  {author} {\bibinfo {author} {\bibfnamefont {H.~T.~T.}\
  \bibnamefont {Nguyen}}\ and\ \bibinfo {author} {\bibfnamefont {H.~M.}\
  \bibnamefont {Le}},\ }\href {\doibase 10.1021/jp3020386} {\bibfield
  {journal} {\bibinfo  {journal} {J. Phys. Chem. A}\ }\textbf {\bibinfo
  {volume} {116}},\ \bibinfo {pages} {4629} (\bibinfo {year}
  {2012})}\BibitemShut {NoStop}%
\bibitem [{\citenamefont {Le}\ and\ \citenamefont
  {Raff}(2010)}]{Le2010_jp907507z}%
  \BibitemOpen
  \bibfield  {author} {\bibinfo {author} {\bibfnamefont {H.~M.}\ \bibnamefont
  {Le}}\ and\ \bibinfo {author} {\bibfnamefont {L.~M.}\ \bibnamefont {Raff}},\
  }\href {\doibase 10.1021/jp907507z} {\bibfield  {journal} {\bibinfo
  {journal} {J. Phys. Chem. A}\ }\textbf {\bibinfo {volume} {114}},\ \bibinfo
  {pages} {45} (\bibinfo {year} {2010})}\BibitemShut {NoStop}%
\bibitem [{\citenamefont {Darley}\ \emph {et~al.}(2008)\citenamefont {Darley},
  \citenamefont {Handley},\ and\ \citenamefont
  {Popelier}}]{Darley2008_ct800166r}%
  \BibitemOpen
  \bibfield  {author} {\bibinfo {author} {\bibfnamefont {M.~G.}\ \bibnamefont
  {Darley}}, \bibinfo {author} {\bibfnamefont {C.~M.}\ \bibnamefont {Handley}},
  \ and\ \bibinfo {author} {\bibfnamefont {P.~L.~A.}\ \bibnamefont
  {Popelier}},\ }\href {\doibase 10.1021/ct800166r} {\bibfield  {journal}
  {\bibinfo  {journal} {J. Chem. Theory Comput.}\ }\textbf {\bibinfo {volume}
  {4}},\ \bibinfo {pages} {1435} (\bibinfo {year} {2008})}\BibitemShut
  {NoStop}%
\bibitem [{\citenamefont {Cho}\ \emph {et~al.}(2002)\citenamefont {Cho},
  \citenamefont {No},\ and\ \citenamefont {Scheraga}}]{Cho200277}%
  \BibitemOpen
  \bibfield  {author} {\bibinfo {author} {\bibfnamefont {K.-H.}\ \bibnamefont
  {Cho}}, \bibinfo {author} {\bibfnamefont {K.~T.}\ \bibnamefont {No}}, \ and\
  \bibinfo {author} {\bibfnamefont {H.~A.}\ \bibnamefont {Scheraga}},\ }\href
  {\doibase http://dx.doi.org/10.1016/S0022-2860(02)00299-5} {\bibfield
  {journal} {\bibinfo  {journal} {J. Mol. Struct.}\ }\textbf {\bibinfo {volume}
  {641}},\ \bibinfo {pages} {77} (\bibinfo {year} {2002})}\BibitemShut
  {NoStop}%
\bibitem [{\citenamefont {Malshe}\ \emph {et~al.}(2009)\citenamefont {Malshe},
  \citenamefont {Pukrittayakamee}, \citenamefont {Raff}, \citenamefont {Hagan},
  \citenamefont {Bukkapatnam},\ and\ \citenamefont
  {Komanduri}}]{Malshe2009_1.3231686}%
  \BibitemOpen
  \bibfield  {author} {\bibinfo {author} {\bibfnamefont {M.}~\bibnamefont
  {Malshe}}, \bibinfo {author} {\bibfnamefont {A.}~\bibnamefont
  {Pukrittayakamee}}, \bibinfo {author} {\bibfnamefont {L.~M.}\ \bibnamefont
  {Raff}}, \bibinfo {author} {\bibfnamefont {M.}~\bibnamefont {Hagan}},
  \bibinfo {author} {\bibfnamefont {S.}~\bibnamefont {Bukkapatnam}}, \ and\
  \bibinfo {author} {\bibfnamefont {R.}~\bibnamefont {Komanduri}},\ }\href
  {\doibase http://dx.doi.org/10.1063/1.3231686} {\bibfield  {journal}
  {\bibinfo  {journal} {J. Chem. Phys.}\ }\textbf {\bibinfo {volume} {131}},\
  \bibinfo {eid} {124127} (\bibinfo {year} {2009})}\BibitemShut {NoStop}%
\bibitem [{\citenamefont {Manzhos}\ and\ \citenamefont
  {Carrington}(2006)}]{Manzhos2006_1.2336223}%
  \BibitemOpen
  \bibfield  {author} {\bibinfo {author} {\bibfnamefont {S.}~\bibnamefont
  {Manzhos}}\ and\ \bibinfo {author} {\bibfnamefont {T.}~\bibnamefont
  {Carrington}},\ }\href {\doibase http://dx.doi.org/10.1063/1.2336223}
  {\bibfield  {journal} {\bibinfo  {journal} {J. Chem. Phys.}\ }\textbf
  {\bibinfo {volume} {125}},\ \bibinfo {eid} {084109} (\bibinfo {year}
  {2006})}\BibitemShut {NoStop}%
\bibitem [{\citenamefont {Hollebeek}\ \emph {et~al.}(1999)\citenamefont
  {Hollebeek}, \citenamefont {Ho},\ and\ \citenamefont
  {Rabitz}}]{Hollebeek1999.50.1.537}%
  \BibitemOpen
  \bibfield  {author} {\bibinfo {author} {\bibfnamefont {T.}~\bibnamefont
  {Hollebeek}}, \bibinfo {author} {\bibfnamefont {T.-S.}\ \bibnamefont {Ho}}, \
  and\ \bibinfo {author} {\bibfnamefont {H.}~\bibnamefont {Rabitz}},\ }\href
  {\doibase 10.1146/annurev.physchem.50.1.537} {\bibfield  {journal} {\bibinfo
  {journal} {Annu. Rev. Phys. Chem.}\ }\textbf {\bibinfo {volume} {50}},\
  \bibinfo {pages} {537} (\bibinfo {year} {1999})}\BibitemShut {NoStop}%
\bibitem [{\citenamefont {Ho}\ and\ \citenamefont
  {Rabitz}(2003)}]{Ho2003_1.1603219}%
  \BibitemOpen
  \bibfield  {author} {\bibinfo {author} {\bibfnamefont {T.-S.}\ \bibnamefont
  {Ho}}\ and\ \bibinfo {author} {\bibfnamefont {H.}~\bibnamefont {Rabitz}},\
  }\href {\doibase http://dx.doi.org/10.1063/1.1603219} {\bibfield  {journal}
  {\bibinfo  {journal} {J. Chem. Phys.}\ }\textbf {\bibinfo {volume} {119}},\
  \bibinfo {pages} {6433} (\bibinfo {year} {2003})}\BibitemShut {NoStop}%
\bibitem [{\citenamefont {Eshet}\ \emph {et~al.}(2010)\citenamefont {Eshet},
  \citenamefont {Khaliullin}, \citenamefont {K\"uhne}, \citenamefont {Behler},\
  and\ \citenamefont {Parrinello}}]{PhysRevB.81.184107}%
  \BibitemOpen
  \bibfield  {author} {\bibinfo {author} {\bibfnamefont {H.}~\bibnamefont
  {Eshet}}, \bibinfo {author} {\bibfnamefont {R.~Z.}\ \bibnamefont
  {Khaliullin}}, \bibinfo {author} {\bibfnamefont {T.~D.}\ \bibnamefont
  {K\"uhne}}, \bibinfo {author} {\bibfnamefont {J.}~\bibnamefont {Behler}}, \
  and\ \bibinfo {author} {\bibfnamefont {M.}~\bibnamefont {Parrinello}},\
  }\href {\doibase 10.1103/PhysRevB.81.184107} {\bibfield  {journal} {\bibinfo
  {journal} {Phys. Rev. B}\ }\textbf {\bibinfo {volume} {81}},\ \bibinfo
  {pages} {184107} (\bibinfo {year} {2010})}\BibitemShut {NoStop}%
\bibitem [{\citenamefont {Artrith}\ \emph {et~al.}(2011)\citenamefont
  {Artrith}, \citenamefont {Morawietz},\ and\ \citenamefont
  {Behler}}]{PhysRevB.83.153101}%
  \BibitemOpen
  \bibfield  {author} {\bibinfo {author} {\bibfnamefont {N.}~\bibnamefont
  {Artrith}}, \bibinfo {author} {\bibfnamefont {T.}~\bibnamefont {Morawietz}},
  \ and\ \bibinfo {author} {\bibfnamefont {J.}~\bibnamefont {Behler}},\ }\href
  {\doibase 10.1103/PhysRevB.83.153101} {\bibfield  {journal} {\bibinfo
  {journal} {Phys. Rev. B}\ }\textbf {\bibinfo {volume} {83}},\ \bibinfo
  {pages} {153101} (\bibinfo {year} {2011})}\BibitemShut {NoStop}%
\bibitem [{\citenamefont {Artrith}\ and\ \citenamefont
  {Behler}(2012)}]{PhysRevB.85.045439}%
  \BibitemOpen
  \bibfield  {author} {\bibinfo {author} {\bibfnamefont {N.}~\bibnamefont
  {Artrith}}\ and\ \bibinfo {author} {\bibfnamefont {J.}~\bibnamefont
  {Behler}},\ }\href {\doibase 10.1103/PhysRevB.85.045439} {\bibfield
  {journal} {\bibinfo  {journal} {Phys. Rev. B}\ }\textbf {\bibinfo {volume}
  {85}},\ \bibinfo {pages} {045439} (\bibinfo {year} {2012})}\BibitemShut
  {NoStop}%
\bibitem [{\citenamefont {Sosso}\ \emph {et~al.}(2012)\citenamefont {Sosso},
  \citenamefont {Miceli}, \citenamefont {Caravati}, \citenamefont {Behler},\
  and\ \citenamefont {Bernasconi}}]{PhysRevB.85.174103}%
  \BibitemOpen
  \bibfield  {author} {\bibinfo {author} {\bibfnamefont {G.~C.}\ \bibnamefont
  {Sosso}}, \bibinfo {author} {\bibfnamefont {G.}~\bibnamefont {Miceli}},
  \bibinfo {author} {\bibfnamefont {S.}~\bibnamefont {Caravati}}, \bibinfo
  {author} {\bibfnamefont {J.}~\bibnamefont {Behler}}, \ and\ \bibinfo {author}
  {\bibfnamefont {M.}~\bibnamefont {Bernasconi}},\ }\href {\doibase
  10.1103/PhysRevB.85.174103} {\bibfield  {journal} {\bibinfo  {journal} {Phys.
  Rev. B}\ }\textbf {\bibinfo {volume} {85}},\ \bibinfo {pages} {174103}
  (\bibinfo {year} {2012})}\BibitemShut {NoStop}%
\bibitem [{\citenamefont {Behler}(2011{\natexlab{a}})}]{behler2011neural}%
  \BibitemOpen
  \bibfield  {author} {\bibinfo {author} {\bibfnamefont {J.}~\bibnamefont
  {Behler}},\ }\href@noop {} {\bibfield  {journal} {\bibinfo  {journal} {Phys.
  Chem. Chem. Phys.}\ }\textbf {\bibinfo {volume} {13}},\ \bibinfo {pages}
  {17930} (\bibinfo {year} {2011}{\natexlab{a}})}\BibitemShut {NoStop}%
\bibitem [{\citenamefont {Behler}(2014)}]{behler2014representing}%
  \BibitemOpen
  \bibfield  {author} {\bibinfo {author} {\bibfnamefont {J.}~\bibnamefont
  {Behler}},\ }\href@noop {} {\bibfield  {journal} {\bibinfo  {journal} {J.
  Phys.: Condens. Matter}\ }\textbf {\bibinfo {volume} {26}},\ \bibinfo {pages}
  {183001} (\bibinfo {year} {2014})}\BibitemShut {NoStop}%
\bibitem [{\citenamefont {Behler}(2011{\natexlab{b}})}]{behler2011atom}%
  \BibitemOpen
  \bibfield  {author} {\bibinfo {author} {\bibfnamefont {J.}~\bibnamefont
  {Behler}},\ }\href@noop {} {\bibfield  {journal} {\bibinfo  {journal} {J.
  Chem. Phys.}\ }\textbf {\bibinfo {volume} {134}},\ \bibinfo {pages} {074106}
  (\bibinfo {year} {2011}{\natexlab{b}})}\BibitemShut {NoStop}%
\bibitem [{\citenamefont {Jose}\ \emph {et~al.}(2012)\citenamefont {Jose},
  \citenamefont {Artrith},\ and\ \citenamefont
  {Behler}}]{jose2012construction}%
  \BibitemOpen
  \bibfield  {author} {\bibinfo {author} {\bibfnamefont {K.~J.}\ \bibnamefont
  {Jose}}, \bibinfo {author} {\bibfnamefont {N.}~\bibnamefont {Artrith}}, \
  and\ \bibinfo {author} {\bibfnamefont {J.}~\bibnamefont {Behler}},\
  }\href@noop {} {\bibfield  {journal} {\bibinfo  {journal} {J. Chem. Phys.}\
  }\textbf {\bibinfo {volume} {136}},\ \bibinfo {pages} {194111} (\bibinfo
  {year} {2012})}\BibitemShut {NoStop}%
\bibitem [{\citenamefont {Bart{\'o}k}\ \emph {et~al.}(2013)\citenamefont
  {Bart{\'o}k}, \citenamefont {Kondor},\ and\ \citenamefont
  {Cs{\'a}nyi}}]{bartok2013representing}%
  \BibitemOpen
  \bibfield  {author} {\bibinfo {author} {\bibfnamefont {A.~P.}\ \bibnamefont
  {Bart{\'o}k}}, \bibinfo {author} {\bibfnamefont {R.}~\bibnamefont {Kondor}},
  \ and\ \bibinfo {author} {\bibfnamefont {G.}~\bibnamefont {Cs{\'a}nyi}},\
  }\href@noop {} {\bibfield  {journal} {\bibinfo  {journal} {Phys. Rev. B}\
  }\textbf {\bibinfo {volume} {87}},\ \bibinfo {pages} {184115} (\bibinfo
  {year} {2013})}\BibitemShut {NoStop}%
\bibitem [{\citenamefont {Rupp}\ \emph {et~al.}(2012)\citenamefont {Rupp},
  \citenamefont {Tkatchenko}, \citenamefont {M\"uller},\ and\ \citenamefont
  {von Lilienfeld}}]{PhysRevLett.108.058301}%
  \BibitemOpen
  \bibfield  {author} {\bibinfo {author} {\bibfnamefont {M.}~\bibnamefont
  {Rupp}}, \bibinfo {author} {\bibfnamefont {A.}~\bibnamefont {Tkatchenko}},
  \bibinfo {author} {\bibfnamefont {K.-R.}\ \bibnamefont {M\"uller}}, \ and\
  \bibinfo {author} {\bibfnamefont {O.}~\bibnamefont {von Lilienfeld}},\ }\href
  {\doibase 10.1103/PhysRevLett.108.058301} {\bibfield  {journal} {\bibinfo
  {journal} {Phys. Rev. Lett.}\ }\textbf {\bibinfo {volume} {108}},\ \bibinfo
  {pages} {058301} (\bibinfo {year} {2012})}\BibitemShut {NoStop}%
\bibitem [{\citenamefont {Sch\"utt}\ \emph {et~al.}(2014)\citenamefont
  {Sch\"utt}, \citenamefont {Glawe}, \citenamefont {Brockherde}, \citenamefont
  {Sanna}, \citenamefont {M\"uller},\ and\ \citenamefont
  {Gross}}]{PhysRevB.89.205118}%
  \BibitemOpen
  \bibfield  {author} {\bibinfo {author} {\bibfnamefont {K.~T.}\ \bibnamefont
  {Sch\"utt}}, \bibinfo {author} {\bibfnamefont {H.}~\bibnamefont {Glawe}},
  \bibinfo {author} {\bibfnamefont {F.}~\bibnamefont {Brockherde}}, \bibinfo
  {author} {\bibfnamefont {A.}~\bibnamefont {Sanna}}, \bibinfo {author}
  {\bibfnamefont {K.~R.}\ \bibnamefont {M\"uller}}, \ and\ \bibinfo {author}
  {\bibfnamefont {E.~K.~U.}\ \bibnamefont {Gross}},\ }\href {\doibase
  10.1103/PhysRevB.89.205118} {\bibfield  {journal} {\bibinfo  {journal} {Phys.
  Rev. B}\ }\textbf {\bibinfo {volume} {89}},\ \bibinfo {pages} {205118}
  (\bibinfo {year} {2014})}\BibitemShut {NoStop}%
\bibitem [{\citenamefont {von Lilienfeld}\ \emph {et~al.}(2013)\citenamefont
  {von Lilienfeld}, \citenamefont {Rupp},\ and\ \citenamefont
  {Knoll}}]{von2013representation}%
  \BibitemOpen
  \bibfield  {author} {\bibinfo {author} {\bibfnamefont {O.~A.}\ \bibnamefont
  {von Lilienfeld}}, \bibinfo {author} {\bibfnamefont {M.}~\bibnamefont
  {Rupp}}, \ and\ \bibinfo {author} {\bibfnamefont {A.}~\bibnamefont {Knoll}},\
  }\href@noop {} {\bibfield  {journal} {\bibinfo  {journal} {arXiv preprint
  arXiv:1307.2918}\ } (\bibinfo {year} {2013})}\BibitemShut {NoStop}%
\bibitem [{\citenamefont {Tibshirani}(1996)}]{tibshirani1996regression}%
  \BibitemOpen
  \bibfield  {author} {\bibinfo {author} {\bibfnamefont {R.}~\bibnamefont
  {Tibshirani}},\ }\href@noop {} {\bibfield  {journal} {\bibinfo  {journal} {J.
  R. Stat. Soc. B}\ }\textbf {\bibinfo {volume} {58}},\ \bibinfo {pages} {267}
  (\bibinfo {year} {1996})}\BibitemShut {NoStop}%
\bibitem [{\citenamefont {Hastie}\ \emph {et~al.}(2009)\citenamefont {Hastie},
  \citenamefont {Tibshirani},\ and\ \citenamefont {Friedman}}]{hastieelements}%
  \BibitemOpen
  \bibfield  {author} {\bibinfo {author} {\bibfnamefont {T.}~\bibnamefont
  {Hastie}}, \bibinfo {author} {\bibfnamefont {R.}~\bibnamefont {Tibshirani}},
  \ and\ \bibinfo {author} {\bibfnamefont {J.}~\bibnamefont {Friedman}},\
  }\href@noop {} {\emph {\bibinfo {title} {The Elements of Statistical
  Learning}}},\ \bibinfo {edition} {2nd}\ ed.\ (\bibinfo  {publisher}
  {Springer, New York},\ \bibinfo {year} {2009})\BibitemShut {NoStop}%
\bibitem [{\citenamefont {Seko}\ \emph {et~al.}(2014)\citenamefont {Seko},
  \citenamefont {Takahashi},\ and\ \citenamefont
  {Tanaka}}]{PhysRevB.90.024101}%
  \BibitemOpen
  \bibfield  {author} {\bibinfo {author} {\bibfnamefont {A.}~\bibnamefont
  {Seko}}, \bibinfo {author} {\bibfnamefont {A.}~\bibnamefont {Takahashi}}, \
  and\ \bibinfo {author} {\bibfnamefont {I.}~\bibnamefont {Tanaka}},\ }\href
  {\doibase 10.1103/PhysRevB.90.024101} {\bibfield  {journal} {\bibinfo
  {journal} {Phys. Rev. B}\ }\textbf {\bibinfo {volume} {90}},\ \bibinfo
  {pages} {024101} (\bibinfo {year} {2014})}\BibitemShut {NoStop}%
\bibitem [{\citenamefont {Nelson}\ \emph {et~al.}(2013)\citenamefont {Nelson},
  \citenamefont {Hart}, \citenamefont {Zhou},\ and\ \citenamefont
  {Ozoli{\c{n}}{\v{s}}}}]{nelson2013compressive}%
  \BibitemOpen
  \bibfield  {author} {\bibinfo {author} {\bibfnamefont {L.~J.}\ \bibnamefont
  {Nelson}}, \bibinfo {author} {\bibfnamefont {G.~L.~W.}\ \bibnamefont {Hart}},
  \bibinfo {author} {\bibfnamefont {F.}~\bibnamefont {Zhou}}, \ and\ \bibinfo
  {author} {\bibfnamefont {V.}~\bibnamefont {Ozoli{\c{n}}{\v{s}}}},\
  }\href@noop {} {\bibfield  {journal} {\bibinfo  {journal} {Phys. Rev. B}\
  }\textbf {\bibinfo {volume} {87}},\ \bibinfo {pages} {035125} (\bibinfo
  {year} {2013})}\BibitemShut {NoStop}%
\bibitem [{\citenamefont {Zhou}\ \emph {et~al.}(2014)\citenamefont {Zhou},
  \citenamefont {Nielson}, \citenamefont {Xia},\ and\ \citenamefont
  {Ozoli\ifmmode \mbox{\c{n}}\else \c{n}\fi{}\ifmmode~\check{s}\else
  \v{s}\fi{}}}]{PhysRevLett.113.185501}%
  \BibitemOpen
  \bibfield  {author} {\bibinfo {author} {\bibfnamefont {F.}~\bibnamefont
  {Zhou}}, \bibinfo {author} {\bibfnamefont {W.}~\bibnamefont {Nielson}},
  \bibinfo {author} {\bibfnamefont {Y.}~\bibnamefont {Xia}}, \ and\ \bibinfo
  {author} {\bibfnamefont {V.}~\bibnamefont {Ozoli\ifmmode \mbox{\c{n}}\else
  \c{n}\fi{}\ifmmode~\check{s}\else \v{s}\fi{}}},\ }\href {\doibase
  10.1103/PhysRevLett.113.185501} {\bibfield  {journal} {\bibinfo  {journal}
  {Phys. Rev. Lett.}\ }\textbf {\bibinfo {volume} {113}},\ \bibinfo {pages}
  {185501} (\bibinfo {year} {2014})}\BibitemShut {NoStop}%
\bibitem [{\citenamefont {Mizukami}\ \emph {et~al.}(2014)\citenamefont
  {Mizukami}, \citenamefont {Habershon},\ and\ \citenamefont
  {Tew}}]{Mizukami2014_1.4897486}%
  \BibitemOpen
  \bibfield  {author} {\bibinfo {author} {\bibfnamefont {W.}~\bibnamefont
  {Mizukami}}, \bibinfo {author} {\bibfnamefont {S.}~\bibnamefont {Habershon}},
  \ and\ \bibinfo {author} {\bibfnamefont {D.~P.}\ \bibnamefont {Tew}},\ }\href
  {\doibase http://dx.doi.org/10.1063/1.4897486} {\bibfield  {journal}
  {\bibinfo  {journal} {J. Chem. Phys.}\ }\textbf {\bibinfo {volume} {141}},\
  \bibinfo {eid} {144310} (\bibinfo {year} {2014})}\BibitemShut {NoStop}%
\bibitem [{\citenamefont {Zou}\ and\ \citenamefont
  {Hastie}(2005)}]{RSSB:RSSB503}%
  \BibitemOpen
  \bibfield  {author} {\bibinfo {author} {\bibfnamefont {H.}~\bibnamefont
  {Zou}}\ and\ \bibinfo {author} {\bibfnamefont {T.}~\bibnamefont {Hastie}},\
  }\href {\doibase 10.1111/j.1467-9868.2005.00503.x} {\bibfield  {journal}
  {\bibinfo  {journal} {J. R. Stat. Soc. B}\ }\textbf {\bibinfo {volume}
  {67}},\ \bibinfo {pages} {301} (\bibinfo {year} {2005})}\BibitemShut
  {NoStop}%
\bibitem [{\citenamefont {Bl{\"o}chl}(1994)}]{PAW1}%
  \BibitemOpen
  \bibfield  {author} {\bibinfo {author} {\bibfnamefont {P.~E.}\ \bibnamefont
  {Bl{\"o}chl}},\ }\href@noop {} {\bibfield  {journal} {\bibinfo  {journal}
  {Phys. Rev. B}\ }\textbf {\bibinfo {volume} {50}},\ \bibinfo {pages} {17953}
  (\bibinfo {year} {1994})}\BibitemShut {NoStop}%
\bibitem [{\citenamefont {Kresse}\ and\ \citenamefont {Joubert}(1999)}]{PAW2}%
  \BibitemOpen
  \bibfield  {author} {\bibinfo {author} {\bibfnamefont {G.}~\bibnamefont
  {Kresse}}\ and\ \bibinfo {author} {\bibfnamefont {D.}~\bibnamefont
  {Joubert}},\ }\href@noop {} {\bibfield  {journal} {\bibinfo  {journal} {Phys.
  Rev. B}\ }\textbf {\bibinfo {volume} {59}},\ \bibinfo {pages} {1758}
  (\bibinfo {year} {1999})}\BibitemShut {NoStop}%
\bibitem [{\citenamefont {Perdew}\ \emph {et~al.}(1996)\citenamefont {Perdew},
  \citenamefont {Burke},\ and\ \citenamefont {Ernzerhof}}]{GGA:PBE96}%
  \BibitemOpen
  \bibfield  {author} {\bibinfo {author} {\bibfnamefont {J.~P.}\ \bibnamefont
  {Perdew}}, \bibinfo {author} {\bibfnamefont {K.}~\bibnamefont {Burke}}, \
  and\ \bibinfo {author} {\bibfnamefont {M.}~\bibnamefont {Ernzerhof}},\
  }\href@noop {} {\bibfield  {journal} {\bibinfo  {journal} {Phys. Rev. Lett.}\
  }\textbf {\bibinfo {volume} {77}},\ \bibinfo {pages} {3865} (\bibinfo {year}
  {1996})}\BibitemShut {NoStop}%
\bibitem [{\citenamefont {Kresse}\ and\ \citenamefont {Hafner}(1993)}]{VASP1}%
  \BibitemOpen
  \bibfield  {author} {\bibinfo {author} {\bibfnamefont {G.}~\bibnamefont
  {Kresse}}\ and\ \bibinfo {author} {\bibfnamefont {J.}~\bibnamefont
  {Hafner}},\ }\href@noop {} {\bibfield  {journal} {\bibinfo  {journal} {Phys.
  Rev. B}\ }\textbf {\bibinfo {volume} {47}},\ \bibinfo {pages} {558} (\bibinfo
  {year} {1993})}\BibitemShut {NoStop}%
\bibitem [{\citenamefont {Kresse}\ and\ \citenamefont
  {Furthm{\"u}ller}(1996)}]{VASP2}%
  \BibitemOpen
  \bibfield  {author} {\bibinfo {author} {\bibfnamefont {G.}~\bibnamefont
  {Kresse}}\ and\ \bibinfo {author} {\bibfnamefont {J.}~\bibnamefont
  {Furthm{\"u}ller}},\ }\href@noop {} {\bibfield  {journal} {\bibinfo
  {journal} {Phys. Rev. B}\ }\textbf {\bibinfo {volume} {54}},\ \bibinfo
  {pages} {11169} (\bibinfo {year} {1996})}\BibitemShut {NoStop}%
\bibitem [{\citenamefont {Parlinski}\ \emph {et~al.}(1997)\citenamefont
  {Parlinski}, \citenamefont {Li},\ and\ \citenamefont
  {Kawazoe}}]{parlinski1997first}%
  \BibitemOpen
  \bibfield  {author} {\bibinfo {author} {\bibfnamefont {K.}~\bibnamefont
  {Parlinski}}, \bibinfo {author} {\bibfnamefont {Z.-Q.}\ \bibnamefont {Li}}, \
  and\ \bibinfo {author} {\bibfnamefont {Y.}~\bibnamefont {Kawazoe}},\
  }\href@noop {} {\bibfield  {journal} {\bibinfo  {journal} {Phys. Rev. Lett.}\
  }\textbf {\bibinfo {volume} {78}},\ \bibinfo {pages} {4063} (\bibinfo {year}
  {1997})}\BibitemShut {NoStop}%
\bibitem [{\citenamefont {Togo}\ \emph {et~al.}(2008)\citenamefont {Togo},
  \citenamefont {Oba},\ and\ \citenamefont {Tanaka}}]{PhysRevB.78.134106}%
  \BibitemOpen
  \bibfield  {author} {\bibinfo {author} {\bibfnamefont {A.}~\bibnamefont
  {Togo}}, \bibinfo {author} {\bibfnamefont {F.}~\bibnamefont {Oba}}, \ and\
  \bibinfo {author} {\bibfnamefont {I.}~\bibnamefont {Tanaka}},\ }\href
  {\doibase 10.1103/PhysRevB.78.134106} {\bibfield  {journal} {\bibinfo
  {journal} {Phys. Rev. B}\ }\textbf {\bibinfo {volume} {78}},\ \bibinfo
  {pages} {134106} (\bibinfo {year} {2008})}\BibitemShut {NoStop}%
\end{thebibliography}%

\end{document}